\documentclass[12pt,indent]{article}
\usepackage{a4,latexsym,amsmath,amssymb}
\usepackage[latin1]{inputenc}
\usepackage{float}

\usepackage{booktabs,caption}
\usepackage[flushleft]{threeparttable}
\usepackage{rotating}

\usepackage{natbib}
\usepackage{graphics}
\usepackage[english]{babel}
\usepackage{tabularx}
\usepackage{lscape}
\usepackage{multirow}
\usepackage{rotating}
\usepackage{dsfont}
\usepackage{caption}
\usepackage{subcaption}
\usepackage{bbm}
\usepackage[table]{xcolor}
\usepackage{booktabs}
\usepackage{calc}
\usepackage{ifthen}
\usepackage{url}
\usepackage{colortbl}      
\usepackage{tikz}

\usepackage{blindtext}
\usepackage{scrextend}
\usepackage{mathtools}
\usepackage{verbatim}

\usetikzlibrary{shapes,snakes,matrix}

\newenvironment{tabularsmall}
{ \footnotesize \sffamily \tabular } {
\endtabular
\normalfont }

\newcommand{\E}{\operatorname{E}}      
\newcommand{\var}{\operatorname{var}}
\newcommand{\cov}{\operatorname{cov}}

\newcommand{\betab}{{\boldsymbol{\beta}}}

\newcommand{\deltab}{\boldsymbol{\delta}}

\newcommand{\Phib}{{\boldsymbol{\Phi}}}

\newcommand{\xb}{\boldsymbol{x}}

\newcommand{\yb}{\boldsymbol{y}}

\newcommand{\blanco}[1]{}

\def\d{\displaystyle}

\usepackage{natbib}

\usepackage{geometry}
\usepackage{setspace}
\usepackage{tikz}
\usepackage[]{graphicx}
\usepackage[]{color}
\makeatletter
\def\maxwidth{ %
  \ifdim\Gin@nat@width>\linewidth
    \linewidth
  \else
    \Gin@nat@width
  \fi
}
\makeatother

\definecolor{fgcolor}{rgb}{0.345, 0.345, 0.345}

\usepackage{framed}
\makeatletter
 {\par\unskip\endMakeFramed%
 \at@end@of@kframe}
\makeatother

\definecolor{shadecolor}{rgb}{.97, .97, .97}
\definecolor{messagecolor}{rgb}{0, 0, 0}
\definecolor{warningcolor}{rgb}{1, 0, 1}
\definecolor{errorcolor}{rgb}{1, 0, 0}

\usepackage{alltt}
\IfFileExists{upquote.sty}{\usepackage{upquote}}{}

\newtheorem{theorem}{Proposition}[section]

\begin{document}
\bibliographystyle{chicago}
\sloppy

\makeatletter
\renewcommand{\section}{\@startsection{section}{1}{\z@}%
        {-3.5ex \@plus -1ex \@minus -.2ex}%
        {1.5ex \@plus.2ex}%
        {\reset@font\large\sffamily}}
\renewcommand{\subsection}{\@startsection{subsection}{1}{\z@}%
        {-3.25ex \@plus -1ex \@minus -.2ex}%
        {1.1ex \@plus.2ex}%
        {\reset@font\normalsize\sffamily\flushleft}}
\renewcommand{\subsubsection}{\@startsection{subsubsection}{1}{\z@}%
        {-3.25ex \@plus -1ex \@minus -.2ex}%
        {1.1ex \@plus.2ex}%
        {\reset@font\normalsize\sffamily\flushleft}}
\makeatother



\newsavebox{\tempbox}
\newlength{\linelength}
\setlength{\linelength}{\linewidth-10mm} \makeatletter
\renewcommand{\@makecaption}[2]
{
  \renewcommand{\baselinestretch}{1.1} \normalsize\small
  \vspace{5mm}
  \sbox{\tempbox}{#1: #2}
  \ifthenelse{\lengthtest{\wd\tempbox>\linelength}}
  {\noindent\hspace*{4mm}\parbox{\linewidth-10mm}{\sc#1: \sl#2\par}}
  {\begin{center}\sc#1: \sl#2\par\end{center}}
}



\def\R{\mathchoice{ \hbox{${\rm I}\!{\rm R}$} }
                   { \hbox{${\rm I}\!{\rm R}$} }
                   { \hbox{$ \scriptstyle  {\rm I}\!{\rm R}$} }
                   { \hbox{$ \scriptscriptstyle  {\rm I}\!{\rm R}$} }  }

\def\N{\mathchoice{ \hbox{${\rm I}\!{\rm N}$} }
                   { \hbox{${\rm I}\!{\rm N}$} }
                   { \hbox{$ \scriptstyle  {\rm I}\!{\rm N}$} }
                   { \hbox{$ \scriptscriptstyle  {\rm I}\!{\rm N}$} }  }

\def\d{\displaystyle}\def\d{\displaystyle}

\title{Item Response Thresholds Models }
  \author{Gerhard Tutz \\{\small Ludwig-Maximilians-Universit\"{a}t M\"{u}nchen}\\{\small Akademiestra{\ss}e 1, 80799 M\"{u}nchen}}


\maketitle
\begin{abstract} 
\noindent
A comprehensive  class of models is proposed that can be used for continuous, binary, ordered categorical and count type responses. The  difficulty of items is described by difficulty functions, which replace the item difficulty parameters that are typically used in item response models. They crucially determine the response distribution and make the models very flexible with regard to the range of distributions that are covered. 
The model class contains several widely used models as the binary Rasch model and  the graded response model as special cases, allows for simplifications, and offers a distribution free alternative to count type items. 
A major strength of the models is that they can be used for mixed item formats, when different types of items are combined to  measure  abilities or attitudes. It is an immediate consequence of the comprehensive modeling approach that allows that difficulty functions automatically adapt to the response distribution. Basic properties of the model class are shown. Several real data sets are used to illustrate the flexibility of the  models
\end{abstract}


\section{Introduction}
Modern item response theory provides a variety of models for the measurement of abilities, skills  or  attitudes, see, for example, \citet{lord2008statistical}, \citet{VanderLind2016}, \citet{mair2018modern}.
The history of its evolution has been traced back carefully  by \citet{vanlinditr} and \citet{thissen2020intellectual}.

Essential components of item response theory are that items can be located on the same scale as the ability, that the ability is unobserved (latent), and that the latent variable accounts for observed interrelationship among the item responses \citep{thissen2020intellectual}. In addition it is essential that the  responses are random and have to be described by a probabilistic model to explain their distributions \citep{vanlinditr}. 
These features distinguish item response theory from classical test theory, which uses an a priori score on the entire test by assuming an additive decomposition of an observed test score into a true score and a random error.
 
Item response models are  typically tailored to the type of item. For binary items Rasch models and normal-ogive models  are in common use  \citep{rasch1961general,birnbaum1968some}, for ordered models the graded response model \citep{samejima1995acceleration,samejima2016graded}, the partial credit model 
\citep{Masters:82,glas1989extensions} and the sequential model  \citep{Tutz:89f} have been used. For count data items, among others,  Rasch's Poisson count model and extensions as the Conway-Maxwell-Poisson model \citep{forthmann2019revisiting} have been proposed. Continuous response models have been considered by
\citet{samejima1973homogeneous}, \citet{muller1987rasch}, \citet{mellenbergh2016models}. For taxonomies of item response models see \citet{thissen1986taxonomy} and \citet{TuTax2020}.

The threshold model proposed here advances an unifying approach. Rather than developing different models for different types of responses a common response model for all sorts of responses is considered. In the model each item has its own item difficulty function that determines the distribution of the response. Since item difficulty functions  are item-specific the form of the distribution can vary across items. The model class is rather general, it comprises various commonly used models as the binary Rasch model, the normal-ogive model and the graded response model, for the latter it offers a sparser parameterization. It also provides a genuine latent trait model for continuous responses, which can be seen as a latent trait version of classical test theory. In addition to providing a common framework  for existing and novel models it offers a way to combine different types of items in one test, what has been described as mixed item-formats. Instead of using linkage methods \citep{kim2006extension,kolen2014test} to combine different items the model itself accounts for the different sorts of items.

Major advantages of the approach are:
\begin{itemize}
\item The model provides a common framework for several models in common use.
\item A genuine latent trait model as an alternative to classical test theory is contained as a special case. 
\item The model is very flexible and allows for quite different response distributions.
\item Items can have  different formats, they can be continuous, binary or polychotomous, the common model automatically accounts for the distributional differences.
\end{itemize}

The threshold model and basic concepts are introduced in Section \ref{sec:basic}. It is in particular demonstrated how difficulty functions can be used to model the distribution of responses. In Section \ref{sec:discrete} the case of discrete responses is considered and it is shown that common binary models and the graded response model are special cases of threshold models. Section \ref{sec:mixed} is devoted to mixed item formats. In Section \ref{sec:flexible} a more flexible  
way of specifying difficulty functions is given, which allows to let the data determine which function fits best. The computation of estimates is considered  in  Section \ref{sec:inference}, although illustrative applications are given already in previous sections. In the appendix results that are mentioned in the text are given in a more formal way together with proofs. 

\section{Thresholds Models: Basic Concepts}\label{sec:basic}

Let $Y_{pi}$ denote the responses of person $p$ on item $i$ ($p \in \{1,\dots,P\}, i \in \{1,\dots,I\}$) having support $S$. The general thresholds model we propose is given by  
\begin{equation}\label{eq:thr}
P(Y_{pi} > y|\theta_p,\delta_{i}(.))=F(\theta_p-\delta_{i}(y)),
\end{equation} 
where $F(.)$ is a strictly monotonically increasing distribution function, $\theta_p$ is a person parameter and $\delta_{i}(.)$ is a non-decreasing item-specific function, called \textit{item difficulty} function, which is defined on the support $S$. The function $F(.)$ is a \textit{response function}, which to a degree determines the distribution of the response. Since $F(.)$ is increasing for fixed threshold $y$ the probability of a response larger than $y$ increases with increasing person parameter $\theta_p$.
Thus,  $\theta_p$ can be seen as  an ability or attitude parameter, which indicates the tendency of a person to obtain a high score. Higher values of $\theta_p$ are associated with a greater chance of a correct or affirmative response to each item. The name of the model refers to the modeling of the threshold $y$. It  is a threshold on the \textit{observable} scale, not on a latent scale, which are typically considered thresholds in latent trait modeling. 

In addition to the link  between the latent variables and the observable response specified in (\ref{eq:thr}) conditional independence of observable variables given the latent variables is assumed, which is a typical assumption in latent trait theory often referred to as local independence, e.g., \citet{lord1980applications}. Conditional independence together with the latent monotonicity makes the model a monontone latent variable model in the sense of  \citet{holland1986conditional}. Latent monotonicity as defined by \citet{holland1986conditional} means that the probability $P(Y_{pi} > y|\theta_p,\delta_{i}(.))$ is a nondecreasing function of the person parameter for all items. Since the response function $F(.)$ as well as the item difficulty functions are monotone, latent monotonicity holds for the thresholds model.

The specifics of the threshold model follow from the functions that are chosen.
The item difficulty function $\delta_{i}(.)$ contains the properties of the item, in particular if it is easy or hard to score high. It also determines the concrete form of the response distribution, which is only partially determined by $F(.)$. In the following it is shown that the model allows for quite different distributions of responses although the function $F(.)$ is chosen fixed. 


Latent trait models have been extensively discussed for binary or other categorical responses. Nevertheless, we start with the less familiar case of   continuous responses and first investigate  the potential of the thresholds model as a latent trait model for continuous responses.

\subsection{Linear Item Difficulty Functions}
 
A particular interesting item difficulty function is the linear one, which  allows for some simplifications.
Let $Y_{pi}$ be a continuous response variable and the item difficulty be linear, $\delta_{i}(y)= \delta_{0i}+ \delta_i y$, $\delta_i \ge 0$. Then  one obtains that $Y_{pi}$ has distribution function $F(.)$ with the expectation and  variance given by 
\begin{equation}\label{eq:evar}
\E(Y_{pi}) = \gamma_i\theta_p - \gamma_{0i},\quad 
\var(Y_{pi}) = c \gamma_i^2,
\end{equation}
where $\gamma_i=1/\delta_i$, $\gamma_{0i} =(\delta_{0i}+d)/\delta_i$, with constants $d,c$ that are determined by the distribution function $F(.)$, for the concise form of constants  and a proof, see appendix.

It is immediately seen that high ability $\theta_p$ indicates a tendency to high responses. The item parameter $\gamma_i$ is a scaling parameter and $\gamma_{0i}$
is the location on the latent scale. It represents the 'basic' difficulty of the item; if $\gamma_{0i}$ is large, the expected response is small, and vice versa. 
The specific choice $\gamma_i=1$ (equivalent to $\delta_i=1$) yields the simpler forms
$\E(Y_{pi}) = \theta_p - \gamma_{0i},  \var(Y_{pi}) = c$,
which means that the response is simply determined by the difference between ability $\theta_p$ and item difficulty $\gamma_{0i}$, a property that is familiar from the binary Rasch model or the normal-ogive model without a slope parameter.

\subsection{The Person Threshold and the Item Characteristic Function}

The link between the person and the item parameters  can be described and  visualized in several ways. 
An important function is the \textit{person threshold function} (PT function), which for fixed $\theta_p$ is given by
 \[
 g_{i,\theta_p} (y) =P(Y_{pi} > y|\theta_p,\delta_{i}(.)) = F(\theta_p-\delta_{i}(y)),
\]
It shows the probability of an response above $y$ for a specific person with ability $\theta_p$. It is strongly related to the distribution function of $Y_{pi}$, which is simply given by $F_{{pi}}(y) = 1-F(\theta_p-\delta_{i}(y))$. The distribution function is denoted with the subscript ${pi}$ to distinguish it from the response function $F(.)$ (which itself is a distribution function).

The second function is the general 
\textit{item characteristic function} (IC function). It is an extended form of the item characteristic function commonly used in binary item response theory, and  is given by 
 \[
 IC_{i,y} (\theta_p) =P(Y_{pi} > y|\theta_p,\delta_{i}(.)) = F(\theta_p-\delta_{i}(y)).
\]
It shows the  probability for an response above a fixed value $y$ for varying abilities. In contrast to binary models where only the value $y=0$ is interesting, for responses with more than two possible one has more than one function. If the response is continuous  any value $y$ can occur. Thus, the functions depend  on the item $i$ \textit{and} $y$. 

\begin{figure}[H]
\centering
\includegraphics[width=7cm]{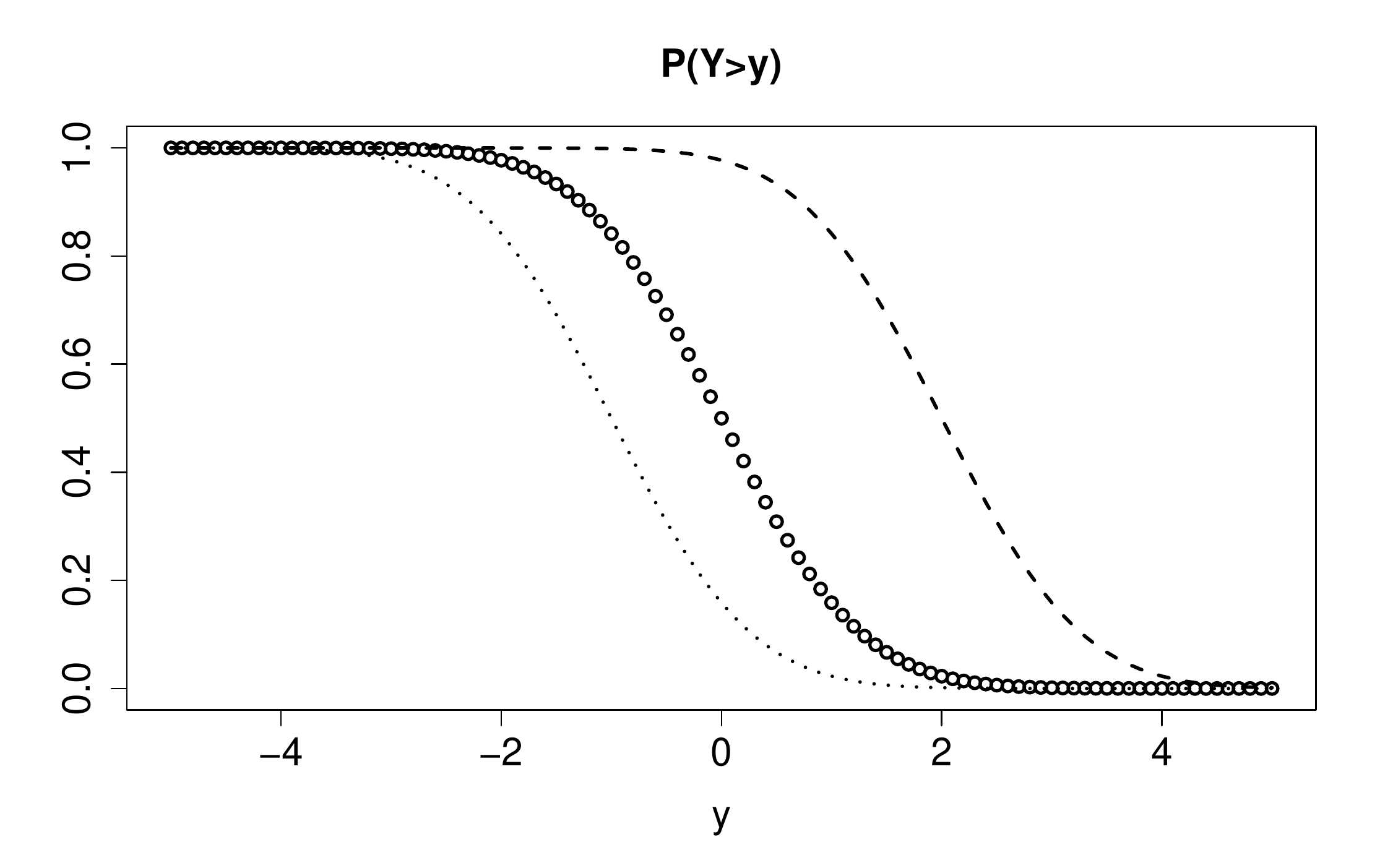}
\includegraphics[width=7cm]{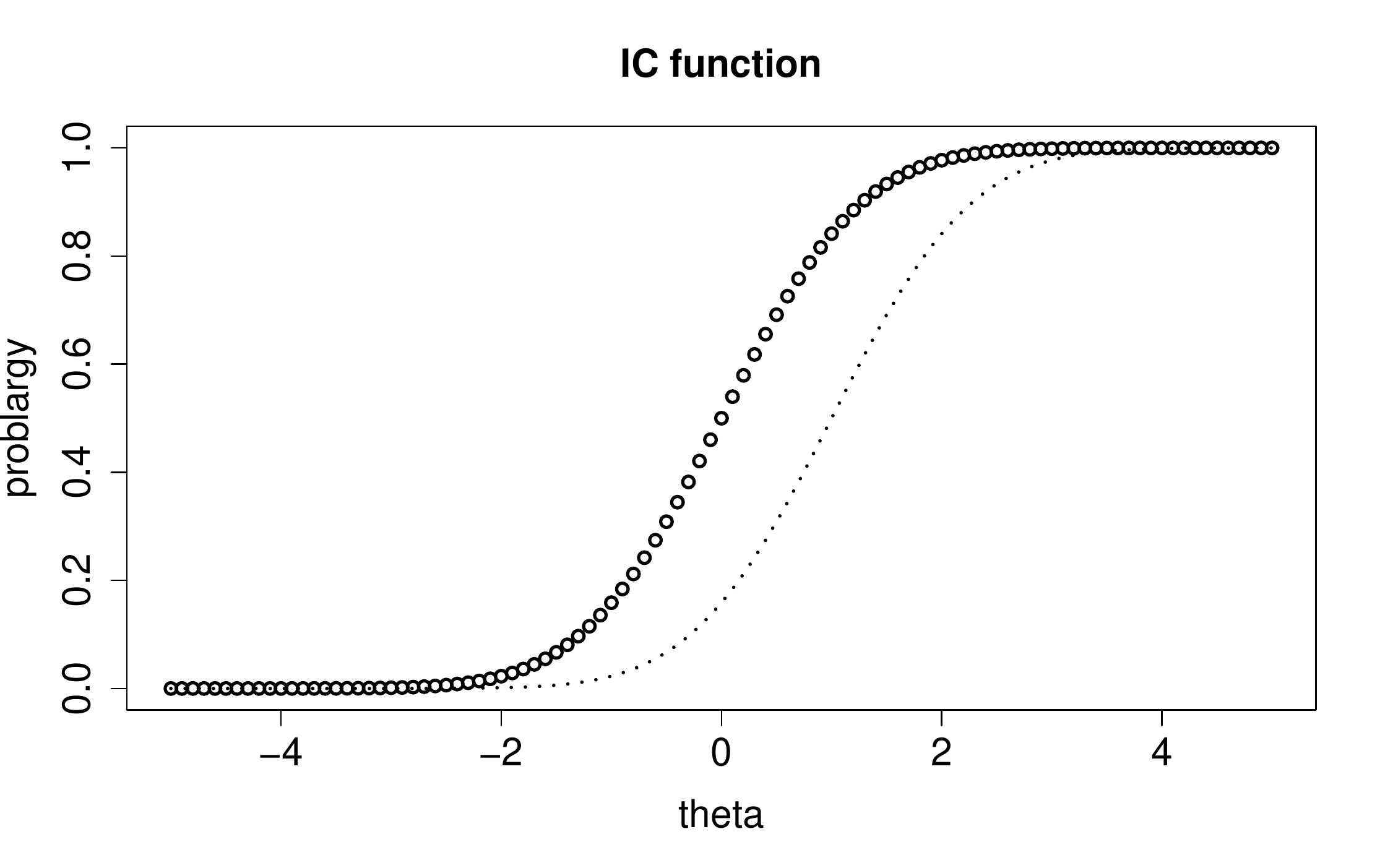}
\caption{Left: Person threshold functions, $P(Y>y)$, for values $\theta=0$ (circles), $\theta=2$  (dashed), $\theta=-1$  (dotted); right: item characteristic functions for $y=0$  (circles) and $y=1$  (dotted)}
\label{fig:curves1}
\end{figure}

For illustration we first consider the simple case of   linear difficulty functions. The left picture in Figure \ref{fig:curves1} shows the person threshold function for three values of $\theta$ if $F(.)$ is the normal distribution function and the  threshold function is linear, $\delta(y)=y$. It is seen that a person with $\theta=2$  (dashed lines) has higher probability of a response above $y$ than a person with $\theta_p=0$ (circles) for all values $y$. The right picture shows the IC function for two values of $y$,  $y=0$  (circles) and $y=1$  (dotted). It is seen that the probability of a response above $y$ is strictly increasing with ability $\theta$. 
In this simple case the PT functions for different $y$ are just shifted versions of the same basic normal distribution function. This changes with the parameters of the   item difficulty function. 

\begin{figure}[H]
\centering
\includegraphics[width=7cm]{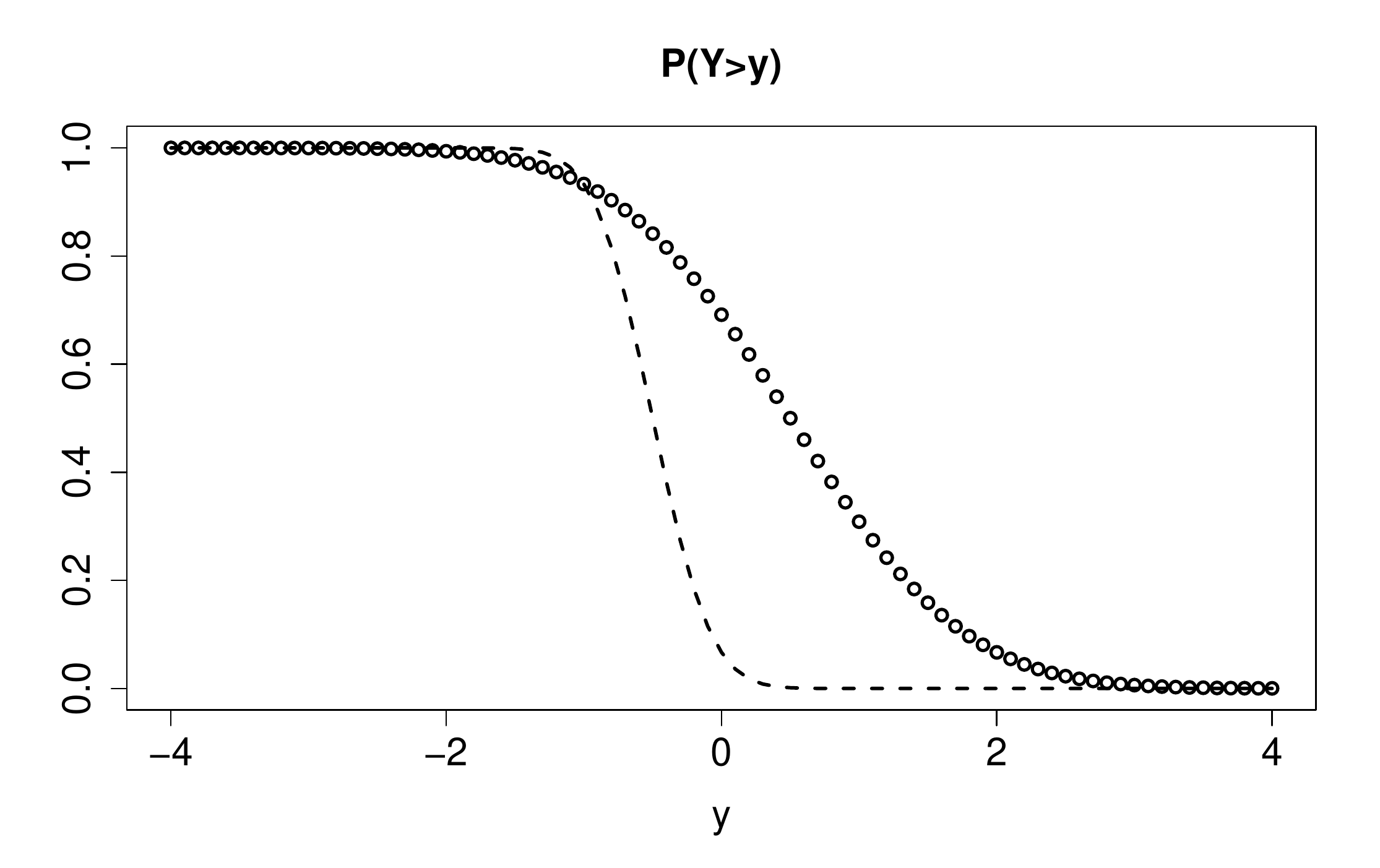}
\includegraphics[width=7cm]{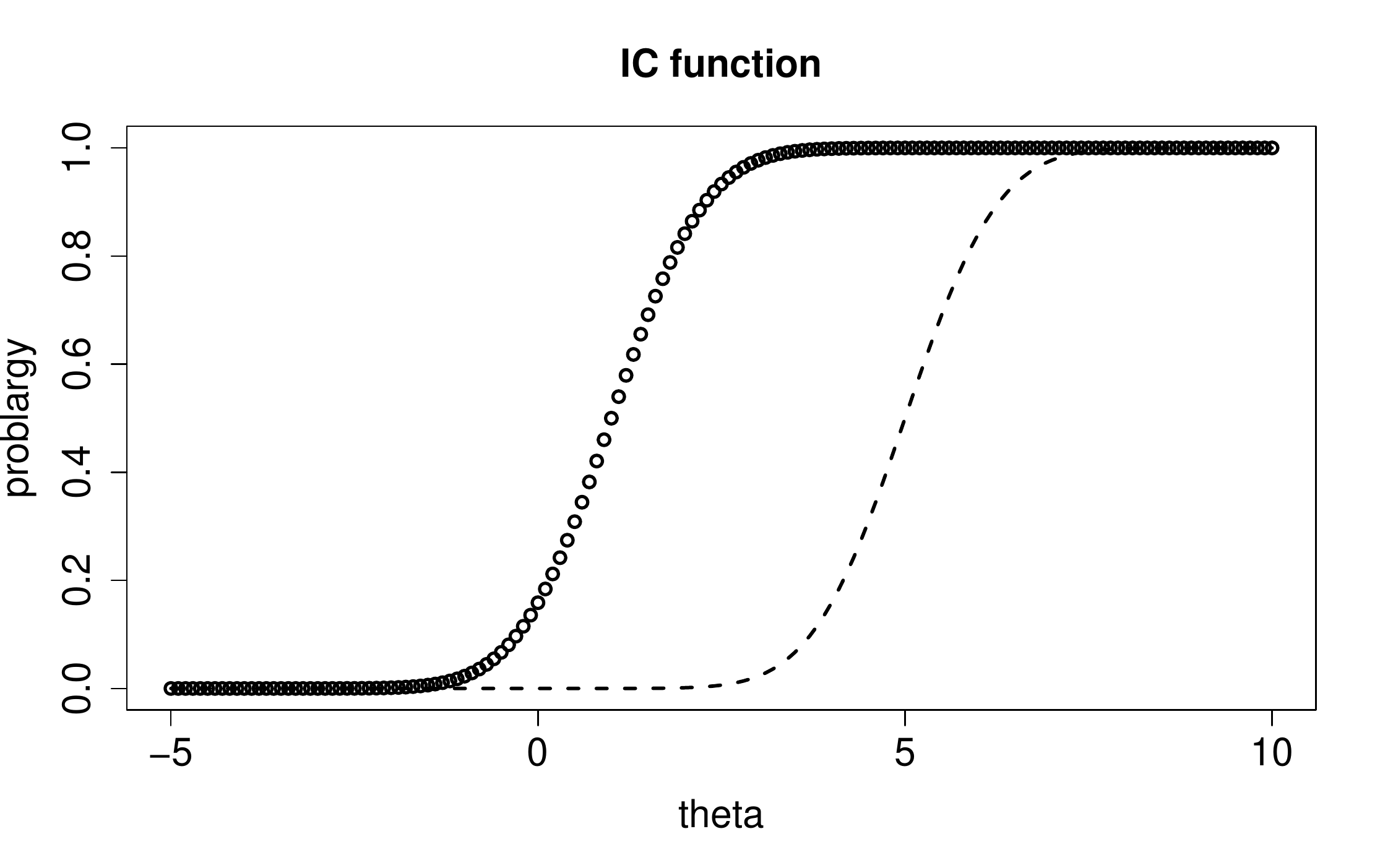}
\caption{Left: Person threshold functions, $P(Y>y)$, for value $\theta = 0.5$ and $\delta_{i}(y)=y$ (circles), $\delta_{i}(y)=2+3y$  (dashed); right: item characteristic functions for the two items for $y=1$}
\label{fig:curves2}
\end{figure}

Therefore, let us consider the parameters of the difficulty function in more detail. The first parameter $\delta_{0i}$ in the  difficulty function $\delta_{i}(y)= \delta_{0i}+ \delta_i y$ determines the location of the item. The corresponding mean of $Y_{pi}$ is $-\delta_{0i}/\delta_i$ (for $\theta_p=0$ and symmetric function $F(.)$). Thus, the PT function is shifted to the left for large location parameter $\delta_{0i} >0$, which represents the basic difficulty.
The second parameter determines the variance of $Y_{pi}$, large $\delta_i$ means that the variance is small. The left picture in Figure \ref{fig:curves2}
shows the PT functions for the simple function $\delta_{i}(y)=y$ (circles) and $\delta_{i}(y)=2+3y$  (dashed). It is seen that for the latter difficulty  function the PT function is shifted to the left and the variance is much smaller, which is seen from the steep decrease of the dashed function. The right picture shows the corresponding item characteristic functions.

It is noteworthy that the IC functions have the same form for all items, namely that of the   distribution function $F(.)$. 
This is immediately seen from the definition of the function $IC_{i,y} (\theta_p) = F(\theta_p-\delta_{i}(y))$ since for fixed value $y$ the value of $\delta_{i}(y)$ is fixed. It is an important aspect regarding interpretation. For any $y$ the item characteristic functions are increasing, and never do cross. That means a person with larger $\theta_p$ than another person has always a larger probability of a response above threshold $y$. This  property, which is well known from binary Rasch models, also holds for the continuous threshold model. It holds in spite of the scaling of the person parameter in the term $\gamma_i\theta_p$ of equ. (\ref{eq:evar}). If the binary Rasch model is extended to the model with a scaling parameter, often referred to as Birnbaum or 2PL model, item characteristic functions typically cross.

The simplicity of the IC functions has an additional  advantage. Since all IC functions are just shifted versions of the same function they show which items are harder, and which are easier to solve. One obtains an ordering without having to investigate the item parameters.

The essential properties of the model, which hold for all sorts of responses to be considered later, can be described by the following functions: 
\begin{itemize} 
\item
The item difficulty function, which characterizes the difficulty of the item over the whole range of possible outcomes.
\item
The person threshold function, which represents the distribution of the responses. The concrete form of the distribution as well as the support (see below) depend on the difficulty functions. The distributions may take quite different forms for different difficulty functions.
\item
The form of the item response function is kept fixed, for all items the probability of scoring above the threshold $y$ increases in the same way with the ability. However, it depends on the threshold $y$ how close items are.
\end{itemize}

It is essential to distinguish between two specifications regarding the complexity of the difficulty function. Let, more general, the difficulty functions be given by  $\delta_{i}(y)= \delta_{0i}+ \delta_i g(y)$, where $g(.)$ is a monotonically increasing function. Then, a simplifying assumption is that the difficulty functions have common slopes, that is, $\delta_1=\dots=\delta_I=\delta$. Without this restriction slopes may vary across items. For linear difficulty functions the assumption of common slopes simply means that for all responses one assumes the same variance, $\var(Y_{pi})$.

\subsection{Links to Classical Test Theory}

The threshold model for continuous responses is a genuine latent trait model. It has all the attributes of a latent trait model,  items are located on the same scale as the latent ability, the latent variable accounts for observed interrelationship among the item responses and  responses are  described by a probabilistic model to explain their distribution. In contrast, classical test theory, which is often used for continuous data, is not a latent trait model in this sense. It is a regression type model, in which an priori score on the entire test is chosen by assuming an additive decomposition of an observed test score into a true score and a random error; it can be traced back to \citet{spearman1961proof}, an extensive presentation is found in  \citet{lord2008statistical}.
 
A similar decomposition is obtained for the thresholds model with linear difficulty functions. If it holds one has
\[
Y_{pi} = E(Y_{pi}) + E_{pi},
\]
where $\E(Y_{pi}) = \gamma_i\theta_p - \gamma_{0i}$, $\var(E_{pi}) = c \gamma_i^2$ and $Y_{pi}$ has distribution function $F(.)$. Random sampling of individuals  yields 
\[
Y_{*i} = E(Y_{*i}) + E_{*i},
\]
which corresponds to the decomposition into  a true-score and an error-score random variable (see Section 2.6 \citet{lord2008statistical}) with the true score depending only on the measurement instrument. The error-score $E_{pi}$   follows distribution function $F(.)$, and has  expectation 0 and variance $c\gamma_{i}^2$. In classical test theory the variance of the error-score is often assumed to be the same for all responses, which means that in addition $\gamma_i =\gamma$ holds for all $i$, or equivalently, that items are homogeneous in the sense that the  slopes do  not vary over items.

Models for continuous responses have also been considered by \citet{mellenbergh2016models}  including Spearman's one factor model and the model for congeneric measurements. The models considered there are, as the classical test theory, rather restrictive since the response  is assumed to  be  a linear function of the latent traits. 

\subsection{Alternative Item Difficulty Functions}

If difficulty functions are linear the responses follow the distribution function $F(.)$. However, responses come with quite different distributions. They can be strictly positive, for example  if the response time is an indicator of the ability of a person, or they are restricted to specific intervals, for example  if a person scores in a given interval continuously or approximately continuously by using numbers, say $1, 2,\dots, 100$. In both cases a normal distribution is inadequate, although in the latter case with numbers $1, 2,\dots, 100$ investigators typically use a normal distribution in spite of the problems that  occur at the boundaries of the interval.

\begin{figure}[h!]
\centering
\includegraphics[width=7cm]{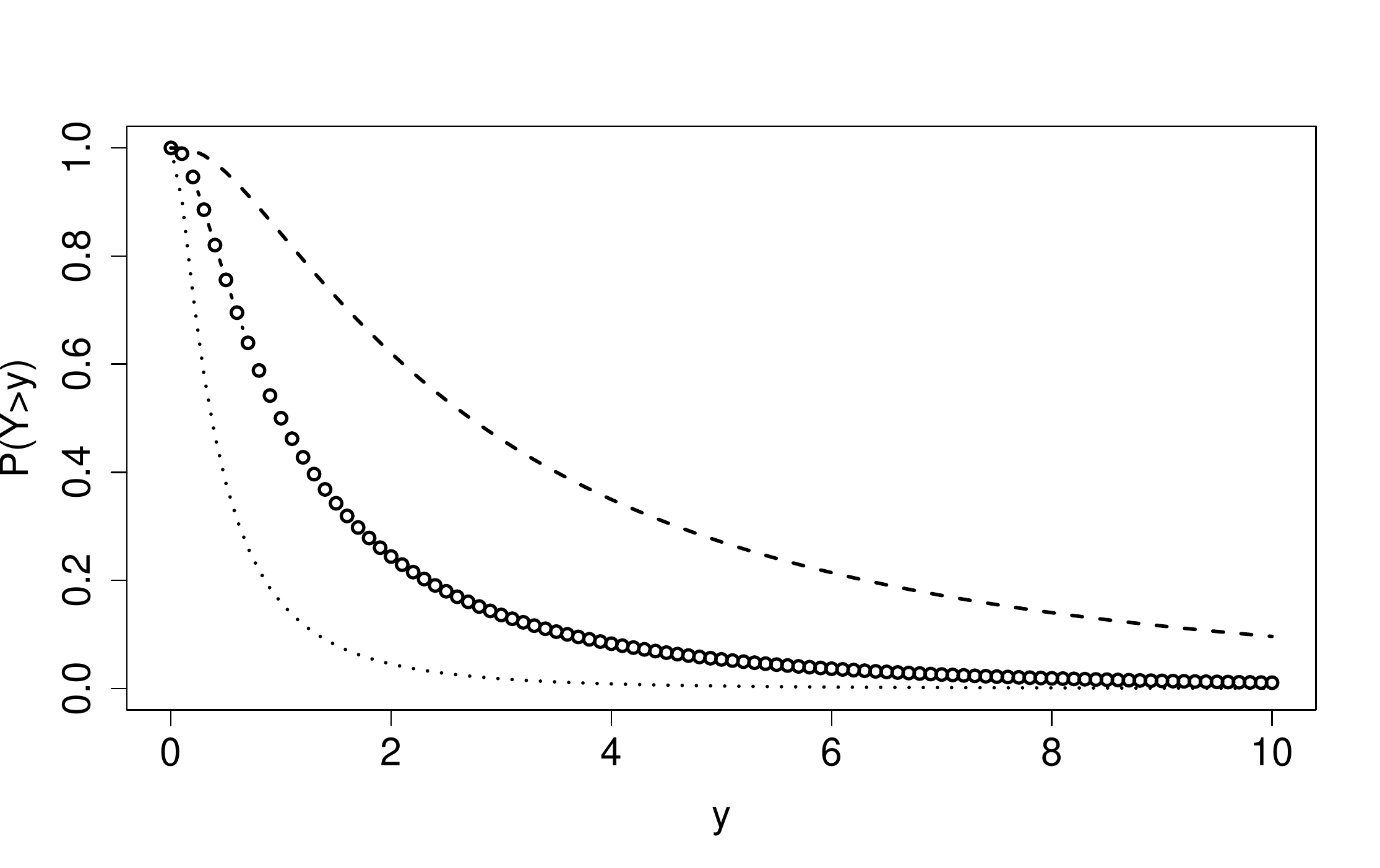}
\includegraphics[width=7cm]{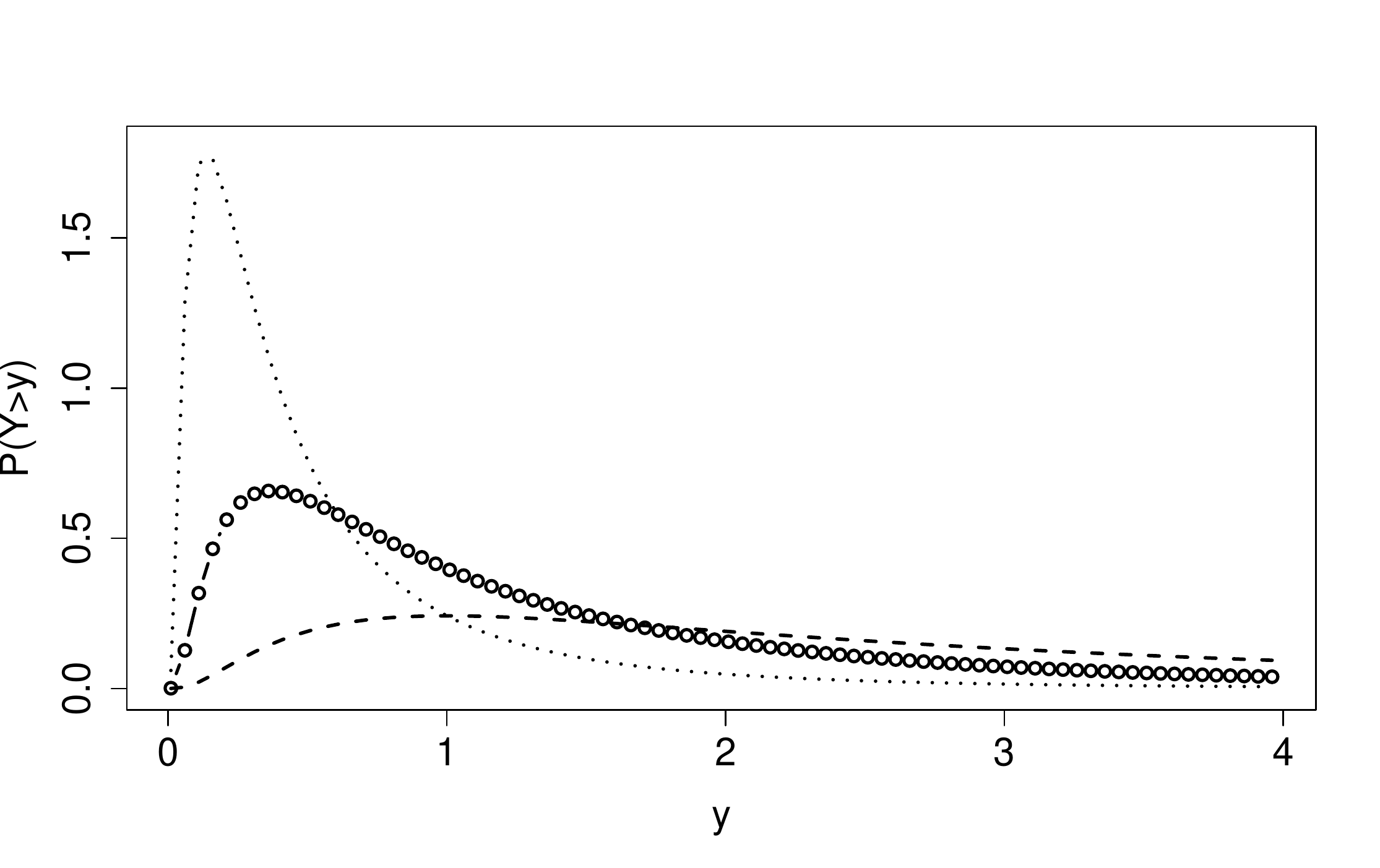}
\includegraphics[width=7cm]{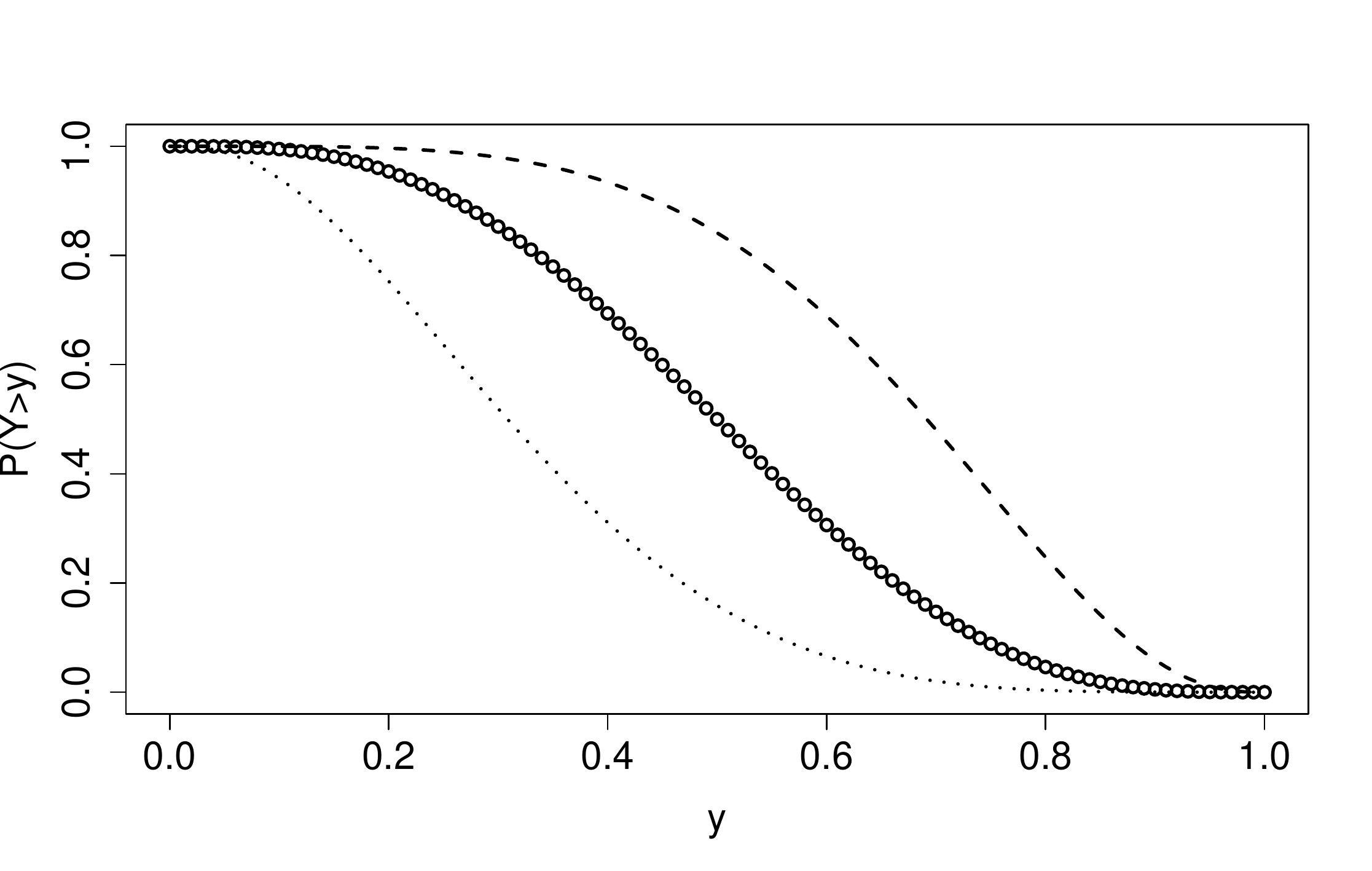}
\includegraphics[width=7cm]{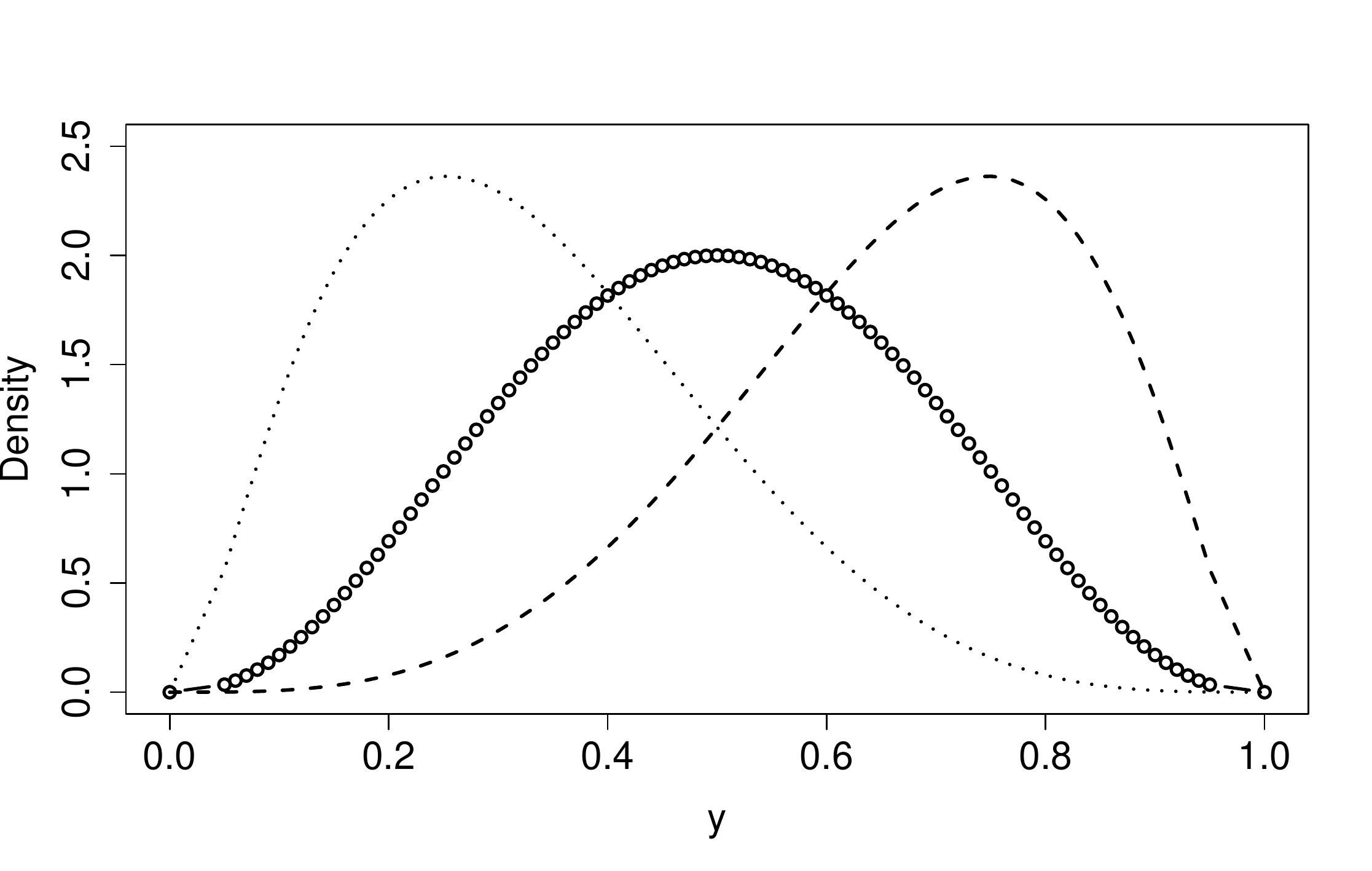}
\caption{Left: $P(Y>y)$ for values $\theta=0$ (circles), $\theta=1$  (dashed), $\theta=-1$  (dotted). Right: corresponding densities }
\label{fig:nonNVcurves}
\end{figure}

A strength of the thresholds model is that it allows to account for the support of the response by using specific difficulty functions. Let the response function $F(.)$ again be the standard normal distribution and the item difficulty be given by $\delta(y)=\log(y)$. Then, one obtains the person threshold functions shown in Figure 
\ref{fig:nonNVcurves} (left, first row) for persons with parameters $\theta=0$ (circles) and $\theta=2$  (dashed). Although a normal distribution is assumed for  the response function  the response is strictly positive, and definitely not normally distributed, as is seen from the corresponding densities (right, first row).

In practice test data are always restricted to specific finite values. A prominent case are Likert-type responses on 5 or 7-point scales. Although values are definitely discrete often they are considered as continuous and common distributions as the normal distribution are assumed. The problem that responses at the boundary can not follow a normal distribution is typically ignored. For truly continuous response scales represented by continuous line segments \citet{samejima1973homogeneous} extended the graded response model to responses to an open line segment, and \citet{muller1987rasch} extended the the rating scale model to responses to a closed line segment. Both extensions are derived as limiting cases of discrete response models.

The threshold model  offers an alternative way to account for the fact that data are restricted to  a fixed interval, and specify a proper distribution for which   the support is the interval in which responses are observed. 
Without loss of generality one can choose the interval $[0,1]$ because data can always be transformed into that interval. Then an attractive difficulty function that can be used is the inverse function  $\delta(y)=aF^{-1}(y)$ with some constant a. The second row of Figure \ref{fig:nonNVcurves} shows the  person threshold functions and the corresponding densities if the response is restricted to the interval $[0,1]$, and $a=1$. It is seen that densities have support $[0,1]$ and are not normally distributed although the normal response function $F(.)$ generates the distribution. For large $\theta_p$ the distribution is shifted to the right, but still within support $[0,1]$. There is no mis-specification of the distribution for very large or small values of $\theta_p$, as occurs if one assumes a normal distribution for the response itself (instead of using a normal response function and appropriate difficulty functions in the thresholds model).

\subsection{Illustrative Application: Cognition Data}
For illustration we use the data set \textit{Lakes} from the R package \textit{MPsychoR} \citep{mair2018modern}. It is a multi-facet G-theory application  taken from \citet{lakes2009applications}. The authors used the  response to assess children's self-regulation in
response to a physically challenging situation. The scale consists of three domains, cognitive, affective/motivational, and physical. We use the cognitive domain only. Each of the 194 children was  rated on six items on his/her self-regulatory ability with  
ratings being on a scale from 1 to 7. 
\citet{mair2018modern} used the data to illustrate concepts of classical test theory implicitly assuming a metric scale level.

\begin{figure}[H]
\centering
\includegraphics[width=7cm]{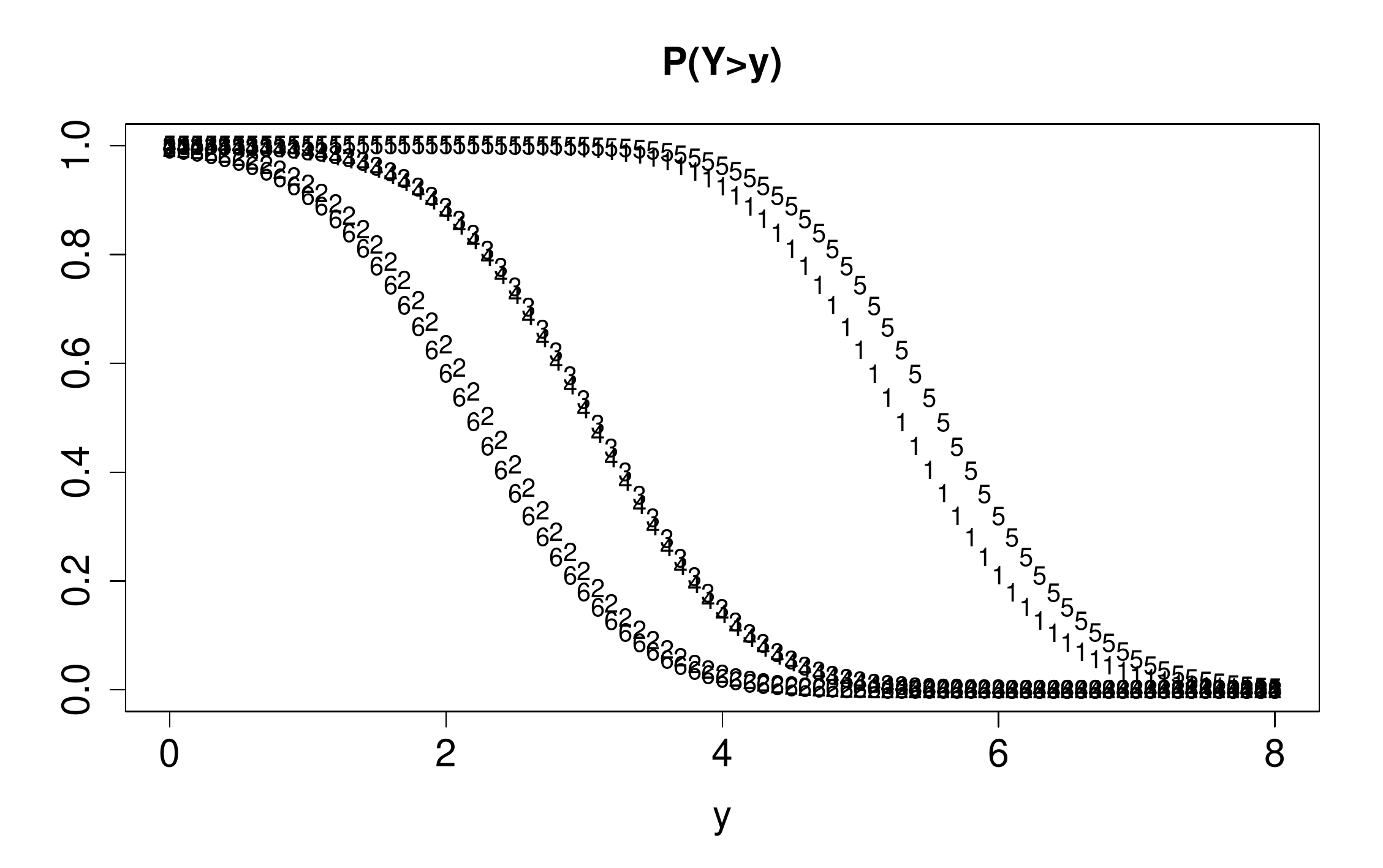}
\includegraphics[width=7cm]{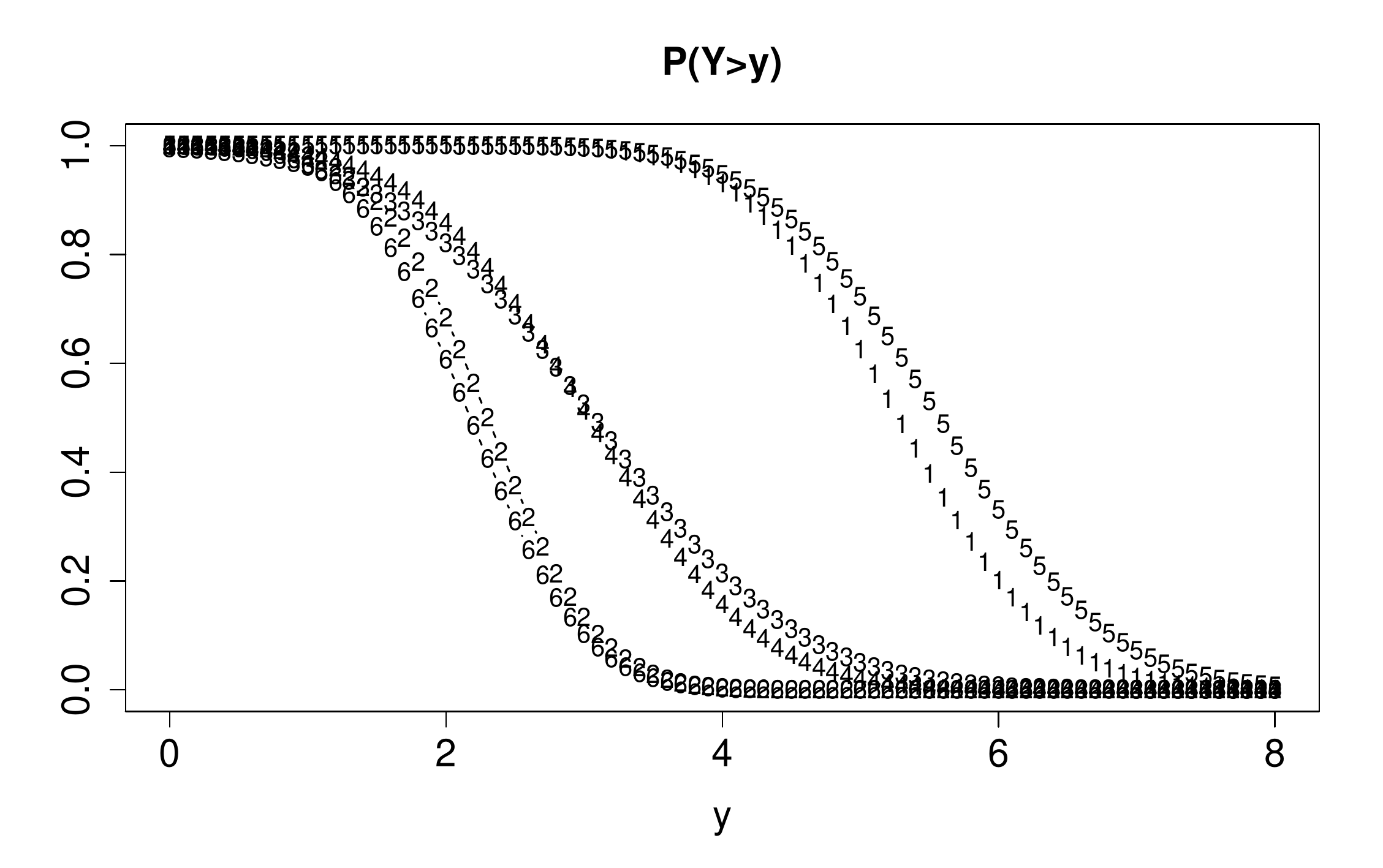}
\includegraphics[width=7cm]{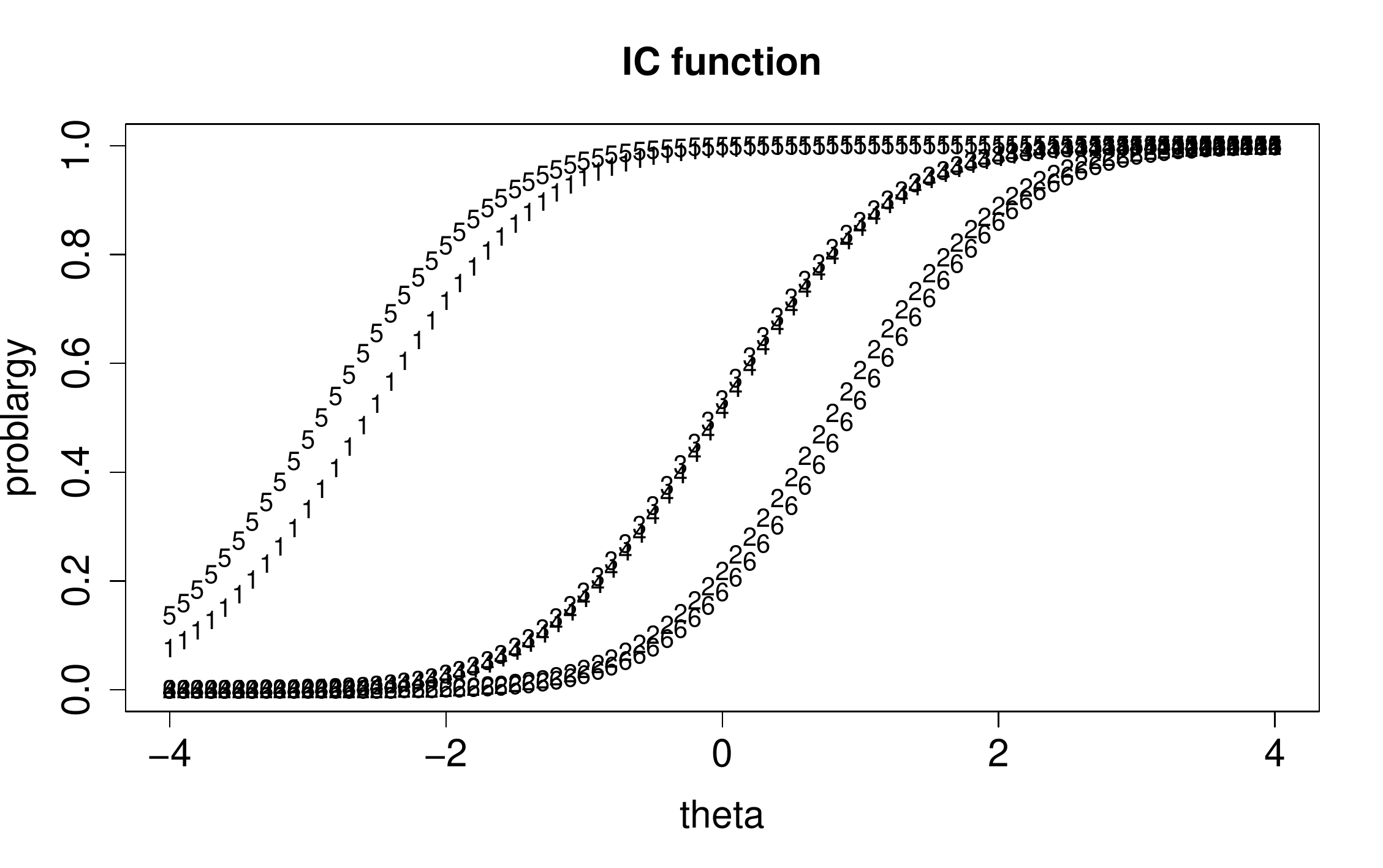}
\includegraphics[width=7cm]{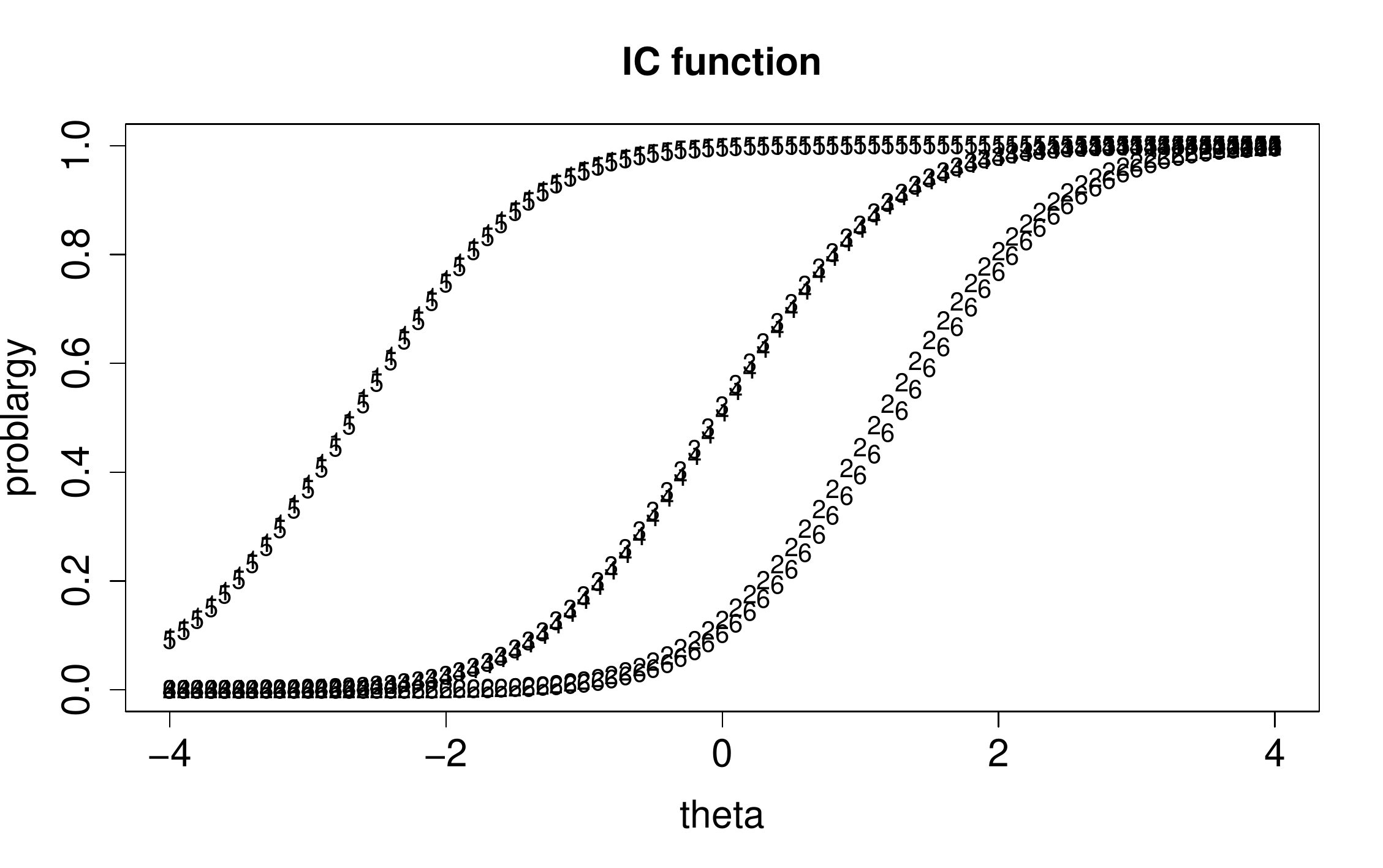}
\includegraphics[width=7cm]{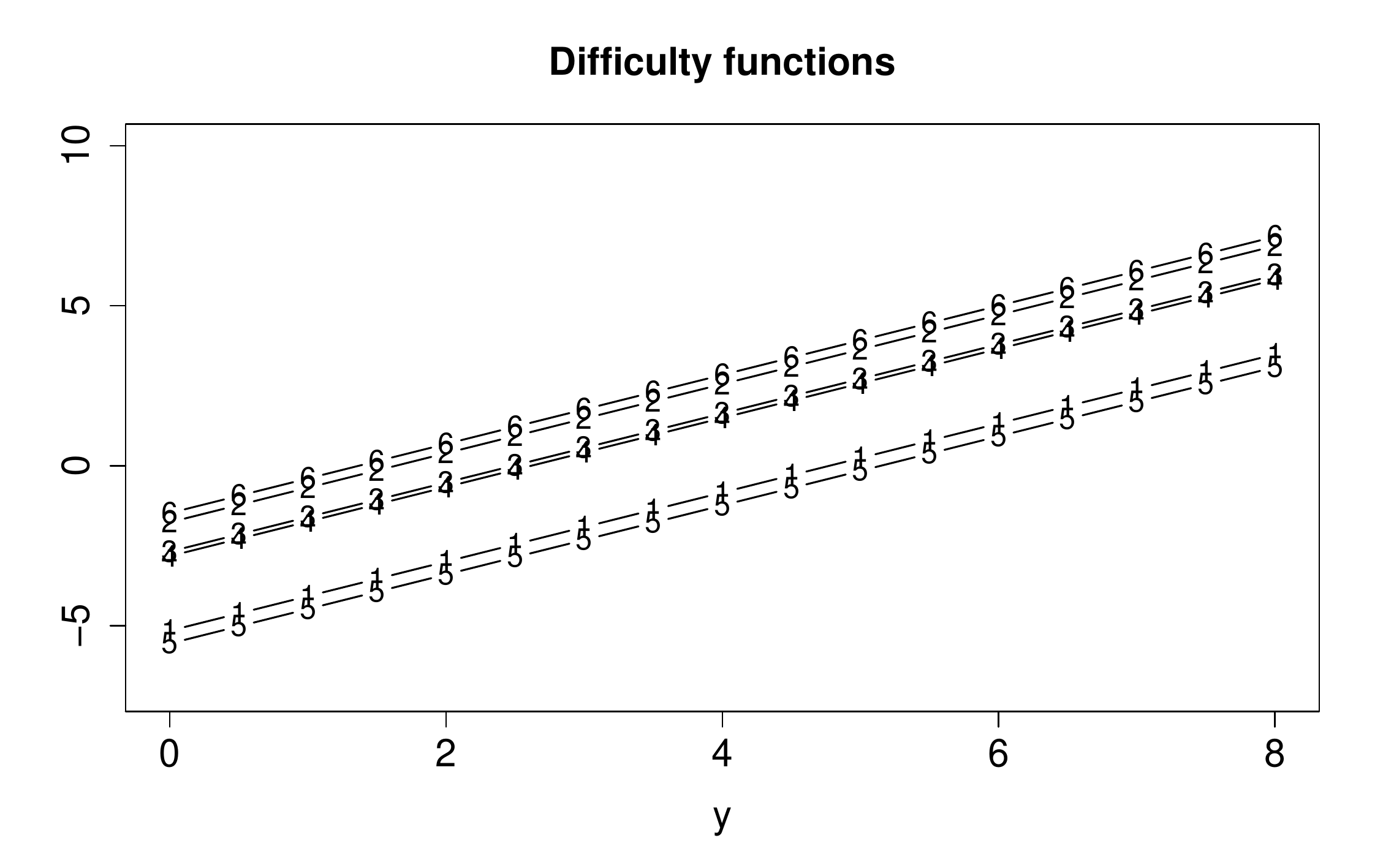}
\includegraphics[width=7cm]{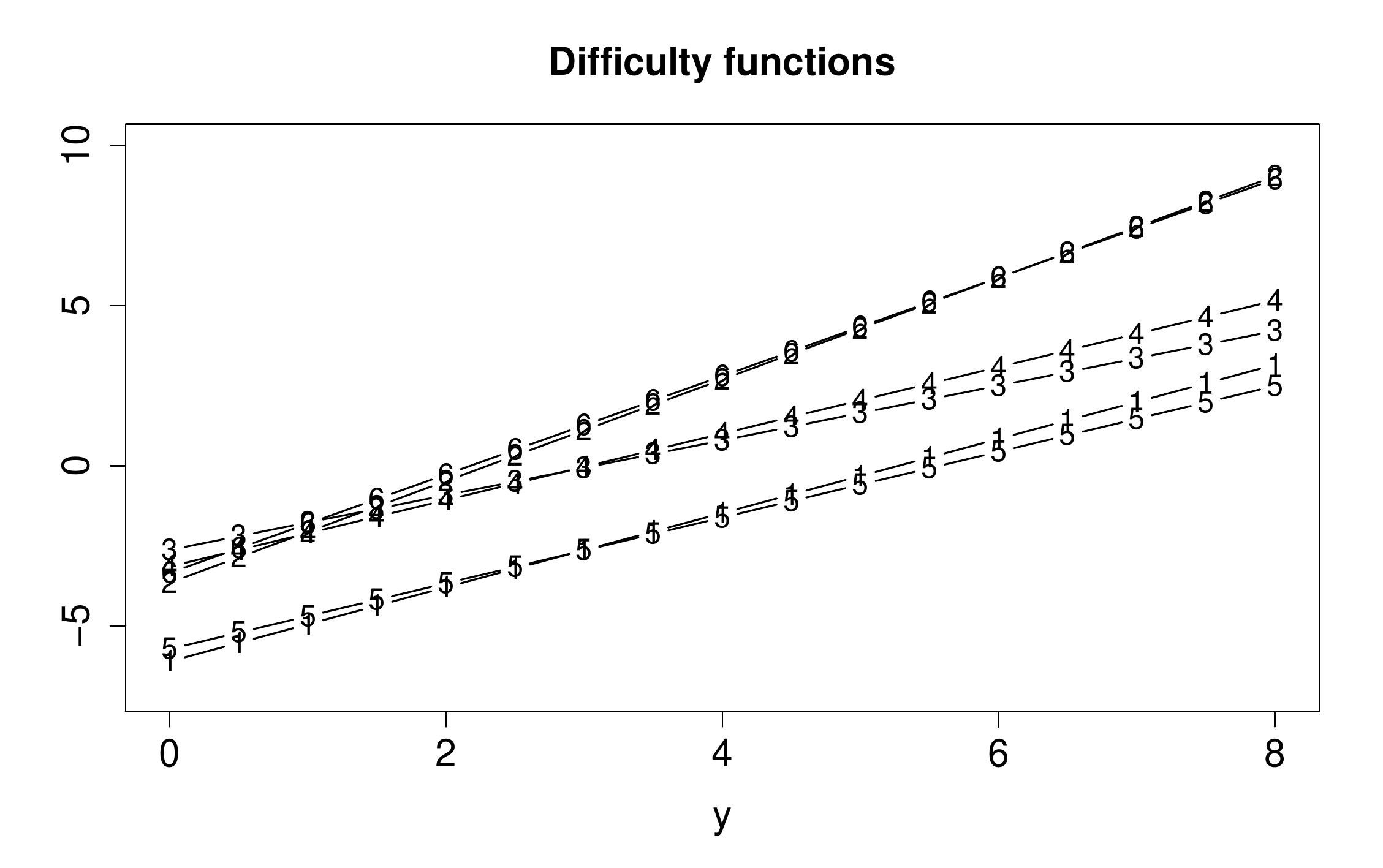}
\caption{First row: person threshold functions, $P(Y>1)$, for cognition data  ($\theta_p = 0$) and linear difficulty functions; second row: IC functions for $y=3$; left: common slopes are assumed, right: varying slopes; third row: difficulty functions.}
\label{fig:lakes1}
\end{figure}

We fit a threshold model with linear difficulty functions and normal response function. The first row of Figure \ref{fig:lakes1} shows the 
person threshold functions for $\theta_p = 0$, under the assumption of common slopes in the difficulty functions (left) and with possibly varying slopes (right).
The numbers in the curves denote the items. It is seen that items 3 and 4 are hardly distinguishable, items 2 and 6 are harder and items 1 and 5 easier. 
It is seen from the right picture (varying slopes) that the variance of responses is smaller for items 2 and 6 when compared to the other items, which corresponds to the large estimated slopes of  items 2 and 6 in Table \ref{tab:cogn3}.
The second row of Figure \ref{fig:lakes1} shows the corresponding IC functions. It is seen that the distance between the pairs of items $\{3,4\}$ and items $\{2,6\}$ is larger if the model allows for varying slopes. The last row shows the difficulty functions. They are strictly parallel in the case of a common slope. For varying slopes the  pairs of items are still close to each other but the items 2 and 6 have larger slopes. 

Since one has nested models it is of interest if the  model with varying slopes can be simplified 
to the model with common slopes in the difficulty functions. The corresponding log-likelihood test is 107.92 on 5 df, which clearly indicates that the simplified model is not adequate.

\begin{table}[H]
 \caption{Estimated parameters for cognition data} \label{tab:cogn3}
\centering
\begin{tabularsmall}{cccccccccc}
  \toprule
 &\multicolumn{1}{c}{ } &\multicolumn{2}{c}{ Common slope } &\multicolumn{2}{c}{ Varying slopes }\\
 \midrule
 &Item  & intercept & slope & intercept & slope \\ 
  \midrule
&1& -5.935405 &1.123863  &-6.109906362 &1.158343367 \\
&2& -2.592028 &1.123863   &3.652171312 &1.587816368\\
&3& -3.452920 &1.123863   &-2.626962524 &0.855081205\\
&4& -3.407838 &1.123863   &-3.147872726 &1.038892679\\
&5& -6.269576 &1.123863   &-5.725666544 &1.027512137\\
&6& -2.451351 &1.123863   &-3.345427435 &1.537712898\\
\midrule
&Log-lik &-1500.006 &&-1446.046\\
\bottomrule
\end{tabularsmall}
\end{table}

\section{Discrete  Responses}\label{sec:discrete}

In the following first it is shown that classical models for binary and ordered responses can be represented as threshold models.
Then models with infinite support are considered.

\subsection{Binary and Ordered Categorical  Responses}

Let us start with the simplest case of a binary response variable $Y_{pi} \in \{0,1\}$. Then, the only relevant value of the function $\delta_{i}(y)$ is $\delta_{i}(0)=\delta_{0i}$ because 
$P(Y_{pi} > 0) = P(Y_{pi} =1)$, and if $P(Y_{pi} =1)$ is known, all response probabilities are known. The thresholds model yields immediately 
the binary response model 
\[
P(Y_{pi} =1|\theta_p,\delta_{i}(.)) = F(\theta_p - \delta_{0i}).
\]
Thus, if $F(.)$ is chosen as normal distribution one obtains the normal-ogive model, if $F(.)$ is the logistic distribution function one obtains the binary Rasch model, \citep{rasch1961general}.

The binary case makes it clear why the difficulty function is defined on the support $S$ of $Y_{pi}$ rather than on the whole field of real numbers.
For $Y_{pi} \in \{0,1\}$ one has to consider only $\delta_{i}(0)$ and $\delta_{i}(1)$. For the latter one has $\delta_{i}(1)=\infty$ since $P(Y_{pi} >1)=1$. For the general case see Proposition \ref{prop:id} in the appendix.

Let now $Y_{pi} \in \{0,\dots, k\}$ be a response variable with ordered categories, and let 
the difficulty functions  $\delta_i(y)$ be restricted only by the assumption that it is a strictly monotonically increasing function. Let parameters be defined by $\delta_{ir}=\delta_i(r-1)$. Then one obtains
the thresholds model 
\[
P(Y_{pi} \ge r|\theta_p,\delta_{i}(.)) = F(\theta_p - \delta_{ir}), \quad r=1,\dots, k,
\]
which is a well known model, namely Samejima's graded response model \citep{samejima1995acceleration,samejima2016graded}.

To obtain the graded response model without further constraints it is essential that the form of the difficulty functions is restricted by the monotonicity assumption only. The monotonicity assumption itself is indispensable because otherwise the  thresholds model would not be defined. 
Nevertheless, it is again interesting to consider the model with a pre-specified threshold  function. If $\delta_{i}(y)= \delta_{0i}+ \delta_i y$ holds one obtains
that differences between adjacent item parameters are constant, $\delta_{ir}-\delta_{i,r-1}=\delta$. 
In this simplified version of the graded response model  each item is characterized by just two parameters,
the location $\delta_{0i}$ and the slope  $\delta_i$. The linear difficulty function assumes some form of equi-distance between categories, which is familiar from other ordered categories models as the Rasch rating scale model \citep{andrich1978rating, Andrichh2016}, which, however is not a graded response model.
Simplified versions also result from using alternative \textit{fixed} difficulty functions, for example, the log function $\delta_{i}(y)= \delta_{0i}+ \delta_i \log(y)$, which has been used above to obtain $Y_{pi} \ge 0$, or the inverse function, which can been used to restrict responses to fixed intervals. 

In particular if the number of  categories is large or medium sized, as for example in a 9-point rating scale,  it is tempting to assume that responses are (approximately) continuous and use corresponding modeling approaches, a strategy that is often found in applied research.  
The graded response takes the support seriously, it is a model that explicitly assumes that the response is discrete and therefore follows a multinomial distribution. The thresholds model, which contains the graded response model as a special case, is quite flexible concerning the assumption of the support. In the general model formulation, $P(Y_{pi} > y|\theta_p,\delta_{i}(.))=F(\theta_p-\delta_{i}(y))$, only the effect of the ability and the item difficulty function on the probability of a response above threshold $y$ is fixed.  It applies to continuous as well as discrete data. Of course, when estimating by maximum likelihood methods one has to distinguish between the discrete and the continuous case since the densities have to be specified. However, in practice the estimated difficulty functions are very similar (see next section), the crucial part is indeed the specification of the response function $F(.)$ and the difficulty function. 

The thresholds model can be seen as bridging the gap between continuous and discrete responses. The bridging can be made more explicit in the case of the graded response function. As shown in the appendix  there is a strong link between the continuous thresholds model and the graded response model since
the thresholds model $P(Y_{pi} > y)=F(\theta_p-\delta_{i}(y))$ holds for continuous response $Y_{pi}$ if and only if the graded response model holds for all categorizations 
\[
Y_{pi}^{(c)} =r \quad \Longleftrightarrow \quad  Y_{pi} \in (\tau_{r}, \tau_{r+1}],
\]
where $\tau_{1} < \dots < \tau_{k}$ are any ordered thresholds. Since the graded response model itself is a thresholds model, this means that  thresholds models are stable under categorization, that is, they also hold if one considers categorized versions of the response. It should be noted that observable responses are considered, the result differs from the usual result that the graded response is a categorized version of a \textit{latent} variable. 
 
\subsection{Political Fears}

As an illustrating example we consider  data from the German Longitudinal Election Study (GLES), which is a long-term study of the German electoral process \citep{GLES}.  The data we are using  originate from the pre-election survey for the German federal  election  in  2017 and are  with political fears. The participants were asked: ``How afraid are you due to the ...'' - (1) refugee crisis?
- (2) global climate change?
- (3) international terrorism?
- (4) globalization?
- (5) use of nuclear energy?
The answers were measured on Likert scales from 1 (not afraid at all) to 7 (very afraid). 
\blanco{
The  explanatory variables in the model are
\textit{Abitur} (high school leaving certificate,  1: Abitur/A levels; 0: else),
\textit{Age} (age of the participant),
\textit{EastWest} (1: East Germany/former GDR; 0: West Germany/former FRG),
\textit{Gender} (1: female; 0: male),
\textit{Unemployment} (1: currently unemployed; 0: else).
 The variable \mbox{EastWest}
refers to the current place of residence where all Berlin residents are assigned to East
Germany.} We  use 200 persons sampled randomly from the available set of observations . 

\begin{figure}[H]
\centering
\includegraphics[width=7cm]{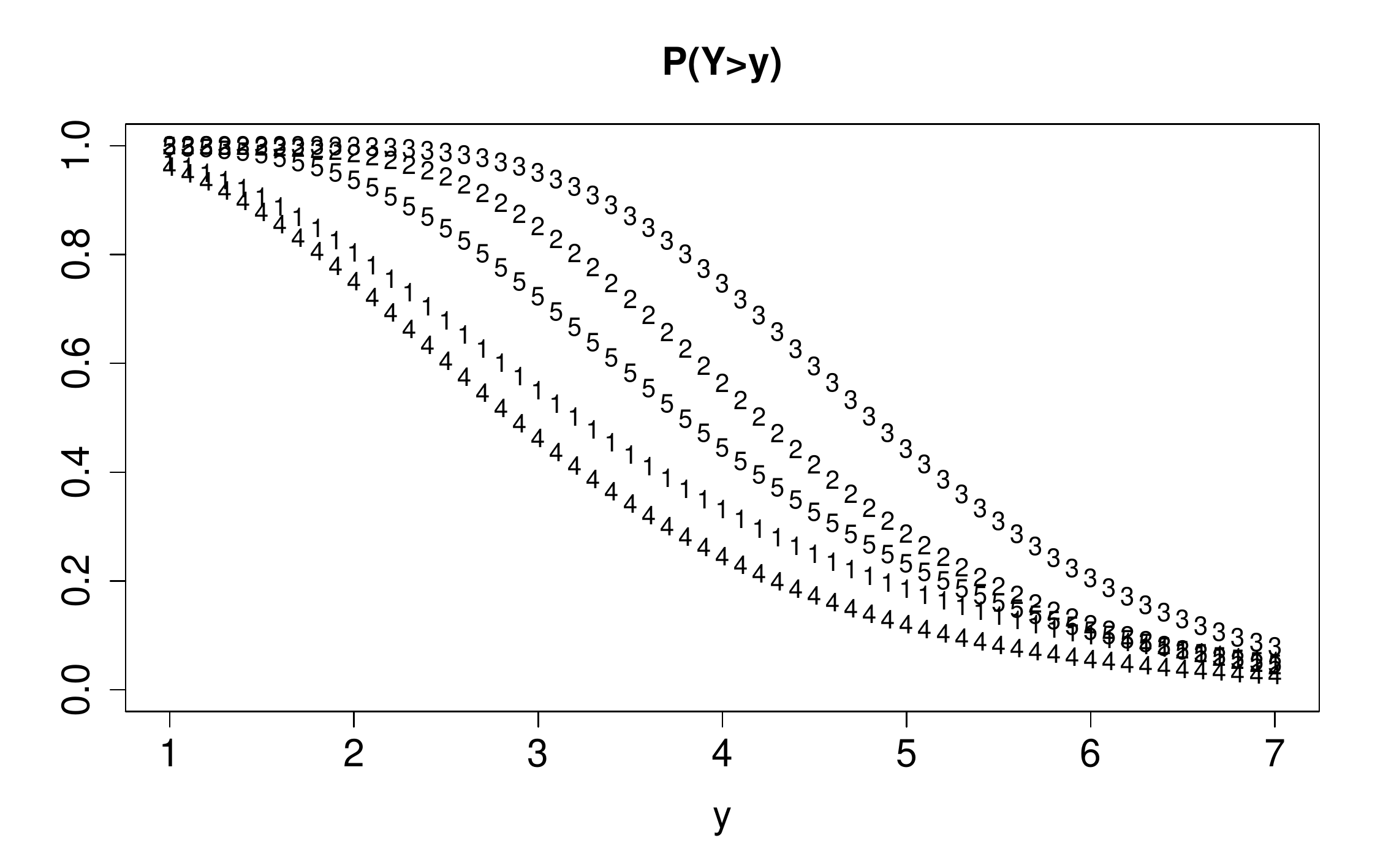}
\includegraphics[width=7cm]{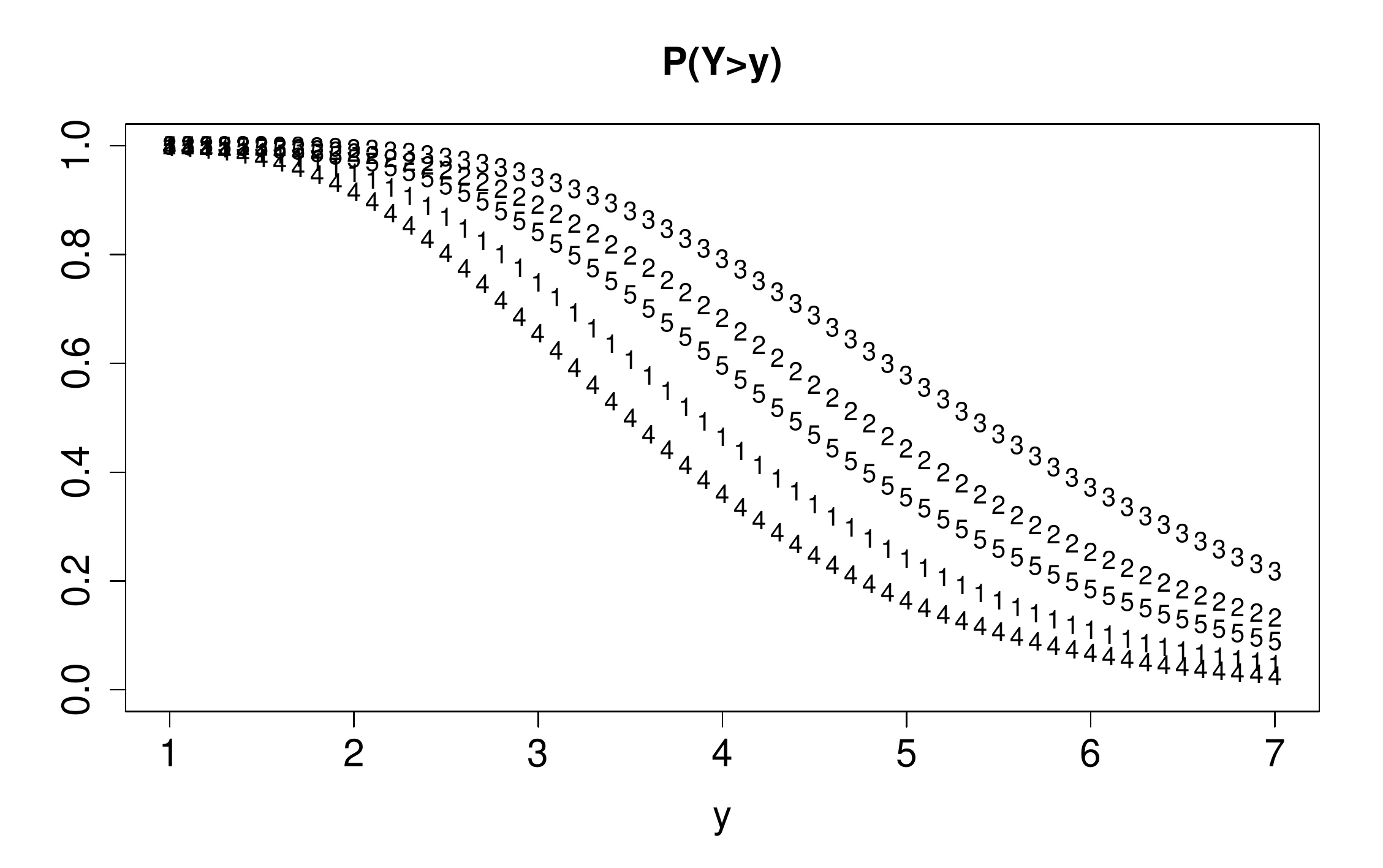}
\caption{PT functions for fear data with logarithmic difficulty functions and common slopes, left: discrete distribution, right: continuous distribution}
\label{fig:fear1}
\end{figure}

Figure \ref{fig:fear1} shows the person threshold functions obtained when using logarithmic difficulty functions (common slopes). The left picture shows the fitted functions when assuming a discrete, multinomial distribution, the right picture when assuming a continuous distribution. It is seen that the fitted person threshold functions are very similar, though log-likelihoods strongly differ (-1987.26 for discrete, -322.28 for continuous distribution).

\subsection{Discrete with Infinite Support: Count Data}

Measurement of cognitive abilities often uses count data, for example, the number of remembered stimuli  \citep{suss2002working},  or the number of generated ideas in a fixed time interval \citep{forthmann2017typing}, for an overview see also \citet{forthmann2019revisiting}. In all these cases the responses are counts with $Y_{pi} \in \{0,1,2,\dots\}$.
A classical model that has been used for this kind of data is Rasch's Poisson count model, which has been extended to the  Conway-Maxwell-Poisson model by
\citet{forthmann2019revisiting}

\begin{figure}[h!]
\centering
\includegraphics[width=7cm]{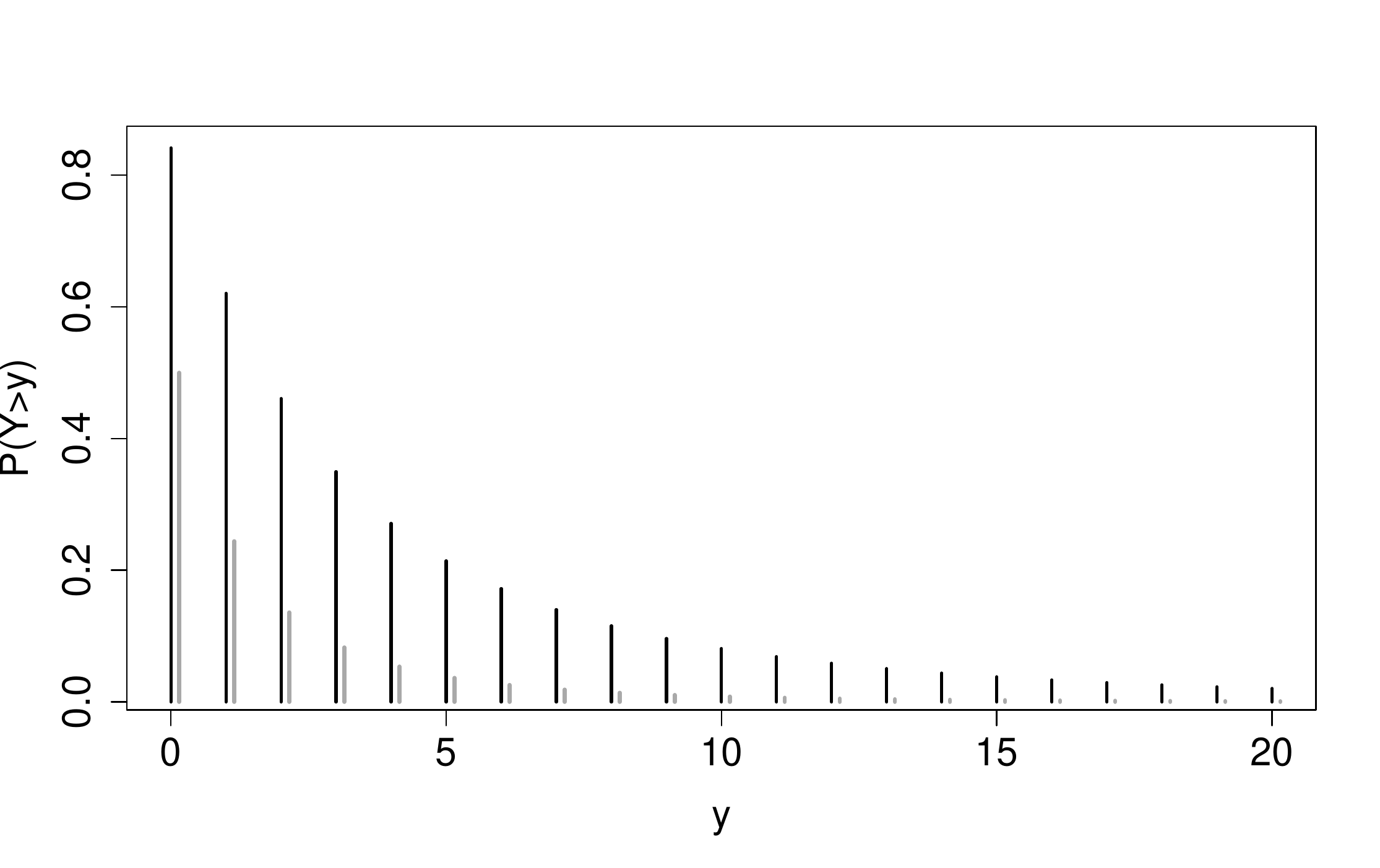}
\includegraphics[width=7cm]{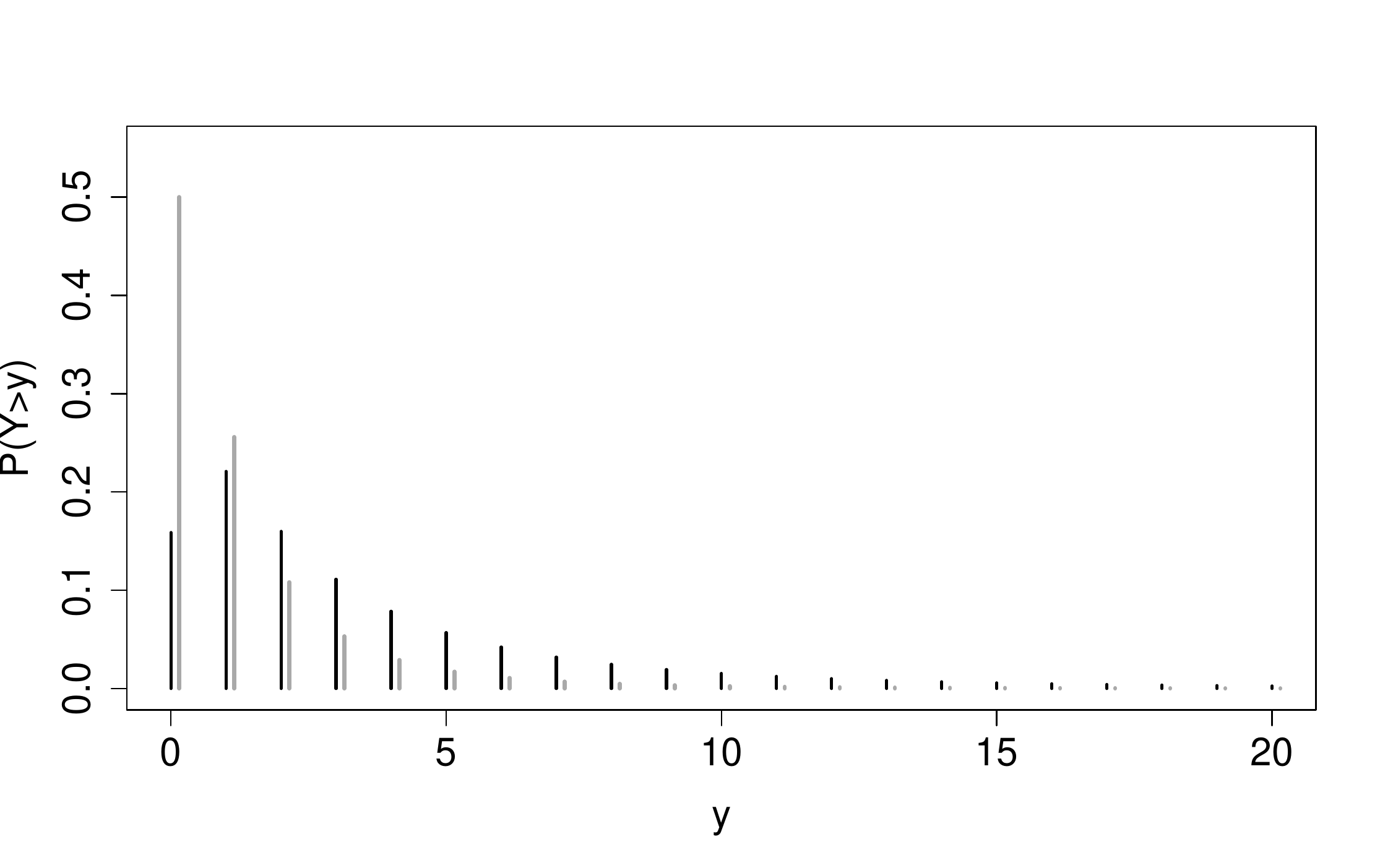}
\caption{ Left: $P(Y>y)$ for values $\theta=1$ (bold),  $\theta=0$  (gray); right: densities}
\label{fig:count1}
\end{figure}

The thresholds model is a flexible alternative to these models. An attractive choice of a fixed difficulty function is the $\log$-function in the form  
$\delta_i(y) = \log(y+1)$. Figure \ref{fig:count1} shows the person threshold functions and the densities for two values of person parameters, $\theta=1$ (bold) and  $\theta=0$  (gray), where $F(.)$ is the standard normal distribution function. It is seen that the PT function for $\theta=1$ is always larger than the 
PT function for $\theta=1$. The densities show that the persons with $\theta=1$ tend to sore higher than persons with $\theta=0$. The IC functions are not shown since by construction they have the form of a normal distribution.

The flexibility of the count thresholds model is comparable to the Conway-Maxwell-Poisson model model if the difficulty functions are specified by 
$\delta_i(y) =\delta_{0i} + \delta_i\log(y+1)$ since the slope $\delta_i$ allows for additional variability  of the response across items.

\subsection{Verbal Fluency Data}
\citet{forthmann2019revisiting} used a data set with four commonly used verbal fluency tasks, which they were so kind to let us use for illustration.  The data
set includes two semantic fluency tasks, namely animal naming (item 1) and naming things that can be found
in a supermarket (item 4) and two letter fluency tasks, words beginning with letter f (item 2) or letter s (item 3).
The 202 participants had one minute to complete each of the verbal fluency tasks. 

\begin{figure}[H]
\centering
\includegraphics[width=7cm]{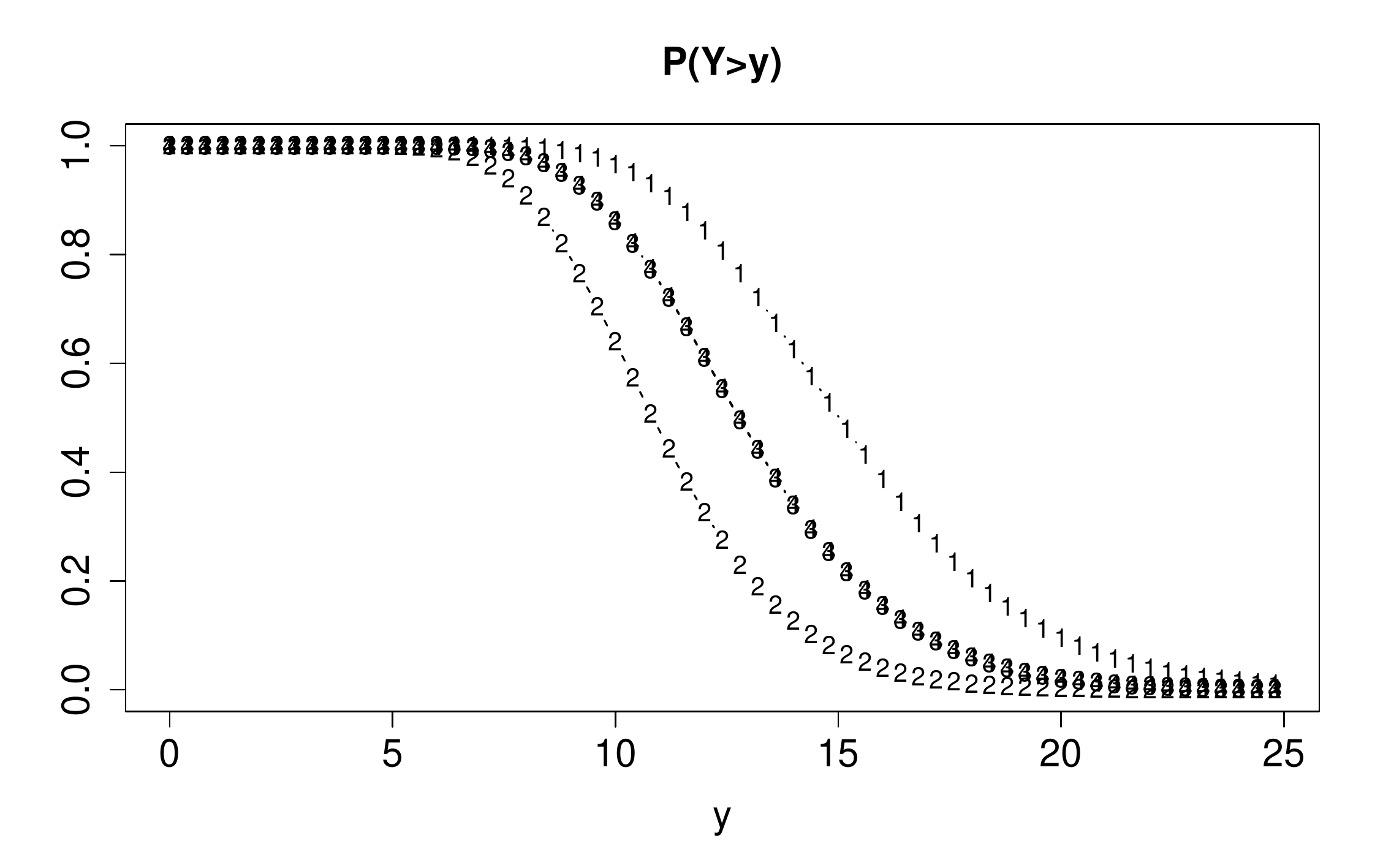}
\includegraphics[width=7cm]{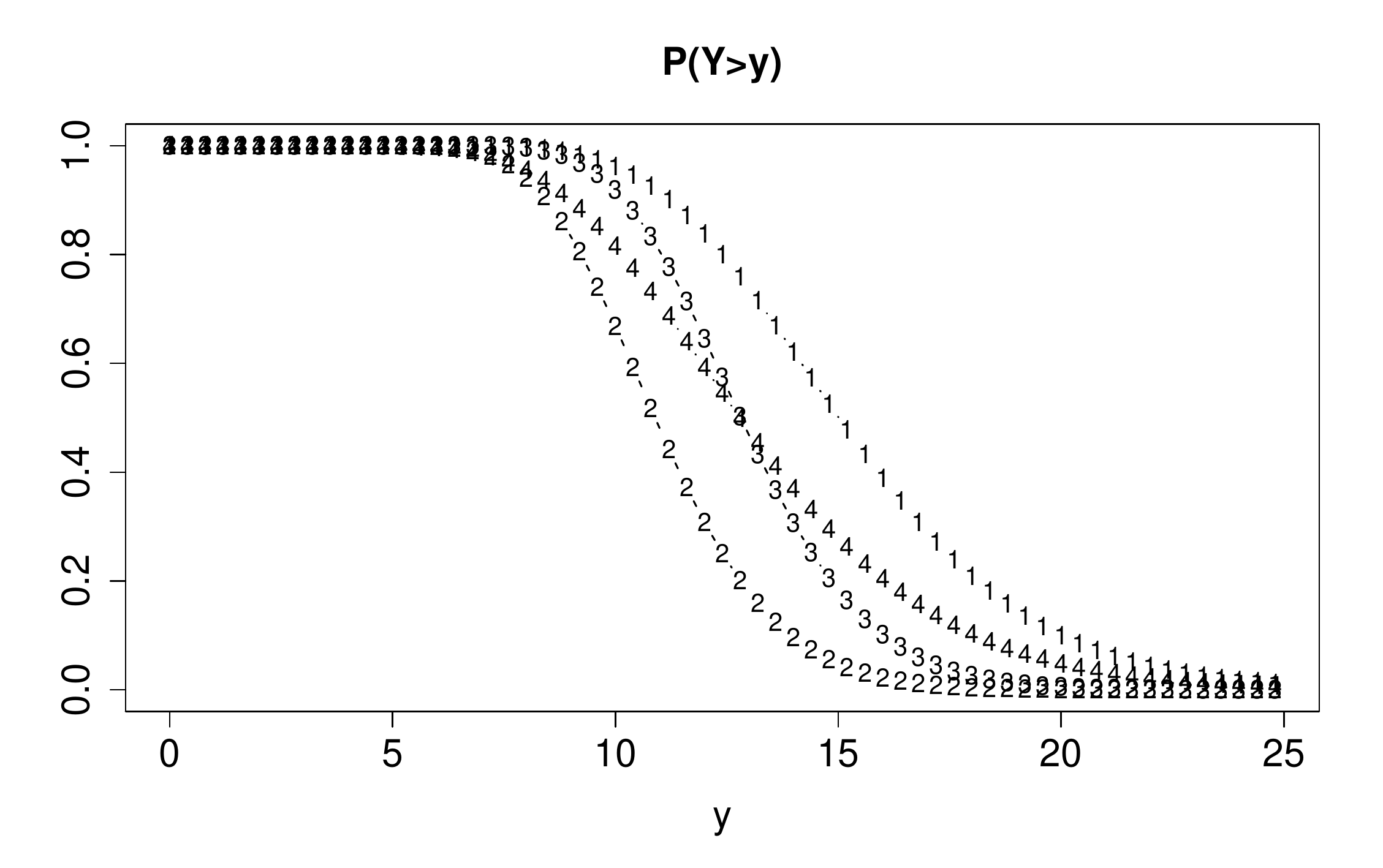}
\caption{Person threshold functions, $P(Y>y)$, for value $\theta = 0$ and $\delta_{i}(y)=log (1+y)$ for verbal fluency data assuming a discrete distribution; left: common slope, right: varying slopes.}
\label{fig:fluency11}
\end{figure}

Figure \ref{fig:fluency11} shows the person threshold functions for common slope (left) and for varying slope (right) if count data are considered as discrete 
(loglik = -2065.42, $\sigma_{\theta}= 1.04$ for common slope, loglik = -2038.58, $\sigma_{\theta}= 1.11$ for varying slopes).
It is seen that under the assumption of a common slope item 3 and 4 have virtually the same threshold function,  item 1 allows for higher responses, item 2 is harder, and counts tend to be lower. If slopes are allowed to vary over items the order of items remains the same but items 3 and 4 have slightly different functions. Item  
3 shows a more distinct decrease indicating smaller dispersion than item 4.

The functions given in Figure \ref{fig:fluency11} are obtained by explicitly using the support $\{0,1,2,\dots\}$, and therefore assuming  a discrete 
distribution. Thus, the curves should be interpreted only at values $\{0,1,2,\dots\}$, only for simplicity of presentation they were shown as continuous functions.  

Since counts are on a metrical scale one could also think of fitting a model that assumes a continuous response, and consider it as an approximation. 
We fitted the corresponding thresholds model and obtained virtually the same functions as given in Figure \ref{fig:fluency11}, which are therefore not shown.
Of course the likelihood values differ from the values obtained by using a discrete model. However,  inference yields similar results. The likelihood ratio test that compares the model with common slopes in the difficulty functions to the model with varying slopes is  53.68 for the discrete model and 54.08 for the continuous model (on 3 df). Thus, in both cases the model with varying slopes turns out to be more appropriate.

\section{Mixed Item Formats}\label{sec:mixed}

Tests often contain a mixture of different item formats. When measuring proficiency  the efficiency of tests can be increased by including binary items, polychotomous items, continuous ones as well as count data items.  The formats of
items in a mixed-format test are often categorized into two classes: multiple choice
(MC) and constructed response (CR). As \citet{kim2006extension} noted, typically, MC items are dichotomously scored
(DS) and CR items are polytomously scored (PS).

There is a considerable body of methods of scale linking for mixed-format tests. The methods are inspired by the linkage methods for data obtained from two groups of examinees through common items \citep{kim2006extension}. Common methods are
mean/mean, mean/sigma,  and Stocking-Lord linkage, see, for example, \citep{hanson2002obtaining,ogasawara2001least,kim2002test,kolen2014test}.

The thresholds model addresses the problem of different item formats in a quite different way. By construction it assumes that there is a common ability that determines the outcome for all items. The model itself does not distinguish between continuous, polychotomous or binary items. 
The only implicit assumption is that it contains order information. Only polychotomous items with un-ordered categories are excluded, but they are less useful  when measuring ability anyway. 

The differences in item formats is captured in the difficulty functions. 
They determine which responses can be expected given a fixed ability parameter, and what distributional form the responses have.  
In the mixed formats case it is not sensible to assume a common slope in the difficulty functions, instead slopes should vary freely, then item difficulty functions automatically  adapt to the item.
Resulting item functions can be quite different for, say, a dichotomous item and an item with five categories. The interpretation has to account for the type of item. For the dichotomous item only the value $\delta(0)$ is relevant while for a five-categories item the set $\delta(j), j=0,\dots,4$ determines the response. For continuous functions the whole difficulty function is interpretable. In contrast to the case of homogeneous items it is less instructive to look at the corresponding item characteristic functions since when considering $P(Y_{pi} > y)$ the value $y$ has quite different meaning for different items.

\begin{figure}[h!]
\centering
\includegraphics[width=7cm]{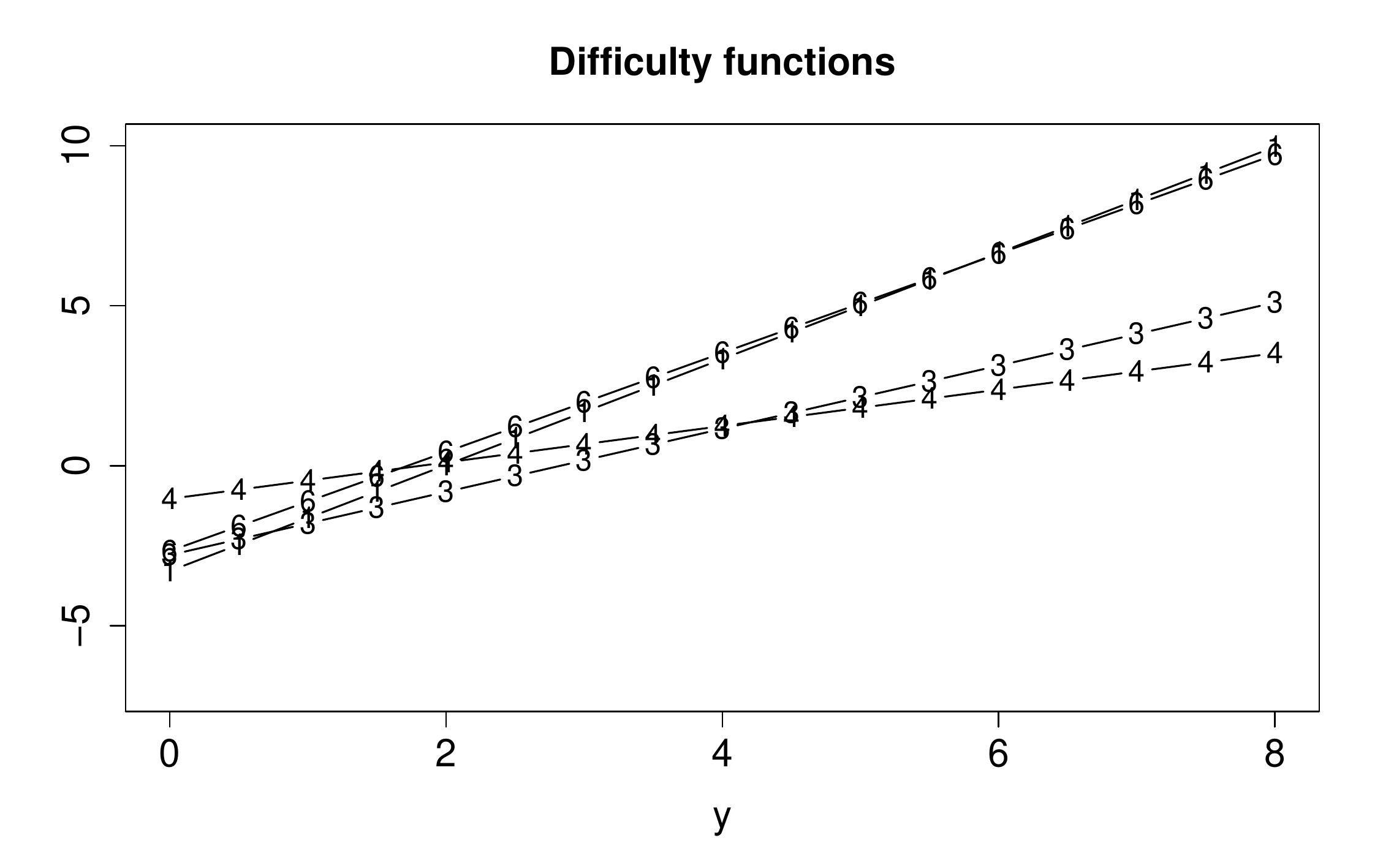}
\includegraphics[width=7cm]{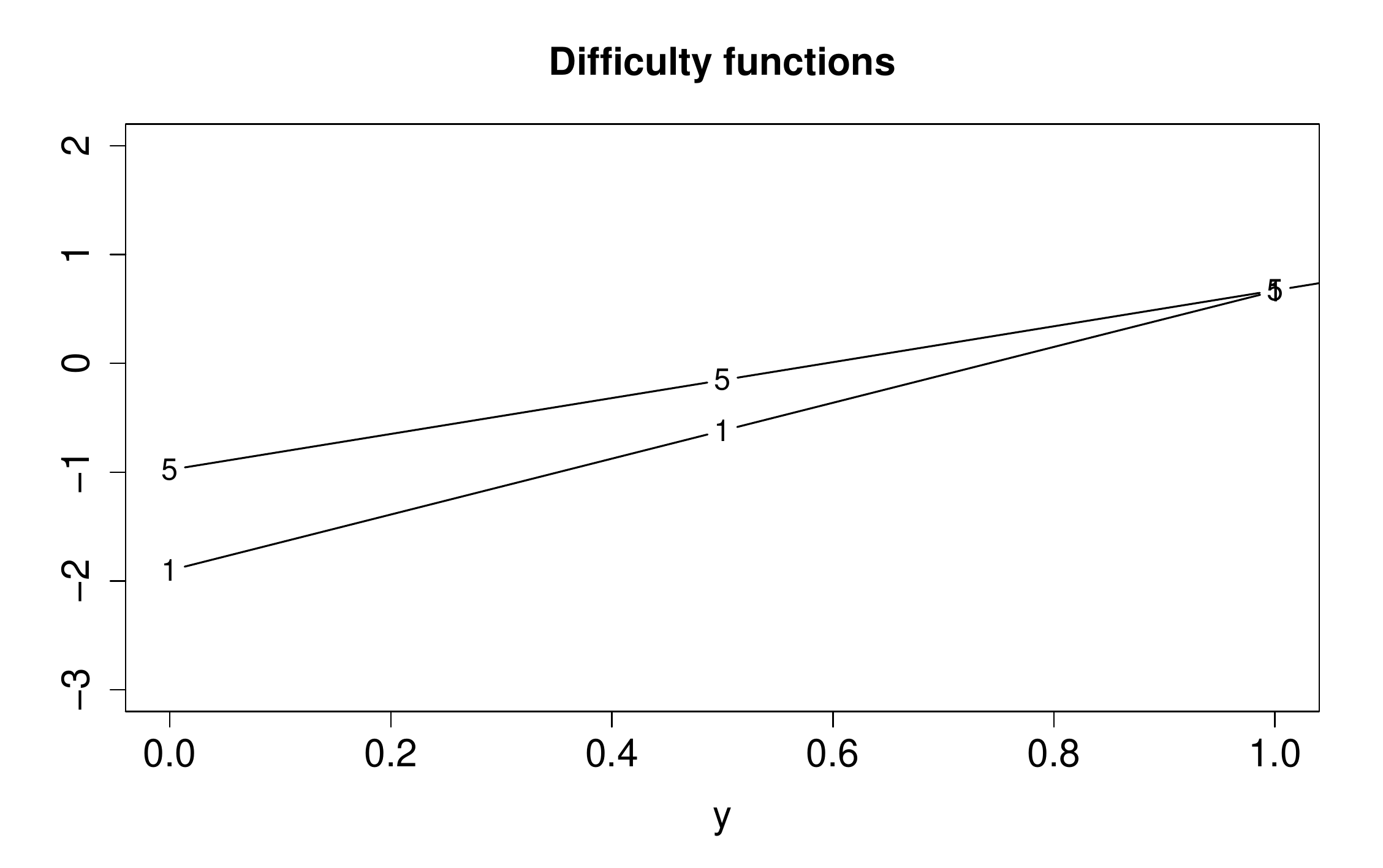}
\includegraphics[width=7cm]{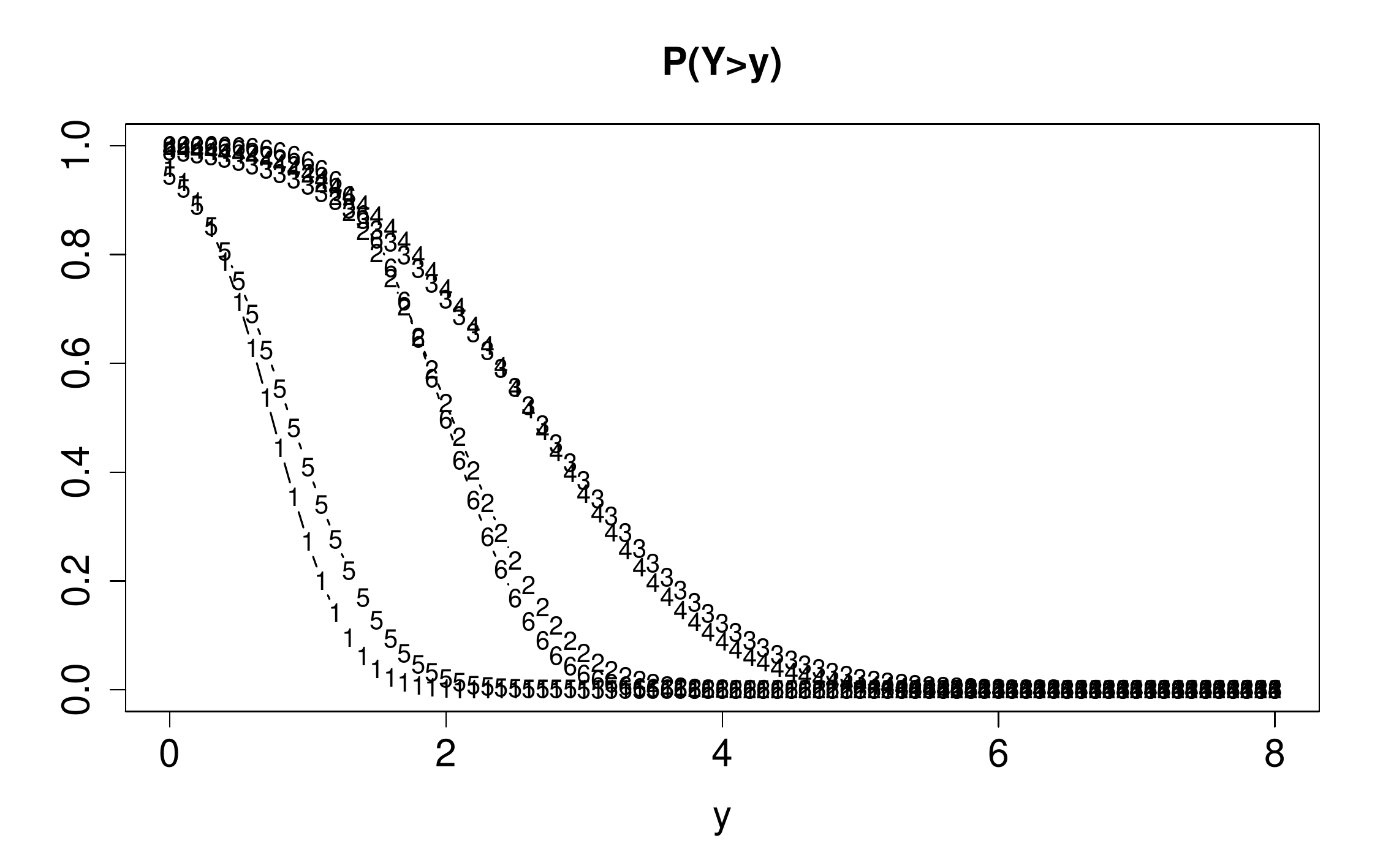}
\caption{Cognition data with  items 1 and 5  as three-categories items. First row: difficulty functions, below: person threshold functions for $\theta=0$.}
\label{fig:lakes2}
\end{figure}

For illustration of   mixed item-formats we consider the cognition data, in which responses range between 1 and 8.  We changed the formats of two  items, item 1 and item 5,  to make them three-categories items by using the thresholds 4 and 6. More precisely, for the items the response is 0 if $Y_{pi} \le 4$, 1 if $ 4 <Y_{pi} \le 6$, and 3 if $  Y_{pi} > 6$. Figure \ref{fig:lakes2} shows the estimated difficulty and PT functions. It is seen that the difficulty functions of the other items (left picture) remain virtually the same as for the original items  shown in Figure \ref{fig:lakes1} (lower right picture). As expected the difficulty functions for items 1 and 5 have changed since now different values of $y$ are relevant. Therefore they are given separately (right picture). 
Figure \ref{fig:lakes2} also shows the corresponding person thresholds functions. Again the curves for item 1 and item 5 are quite different from the curves in Figure \ref{fig:lakes1} because for these items the support is different, namely 0,1,2  (corresponding to 0,4,6 on the original $y$-scale), but for the other items the curves are almost the same.  

\begin{figure}[H]
\centering
\includegraphics[width=7cm]{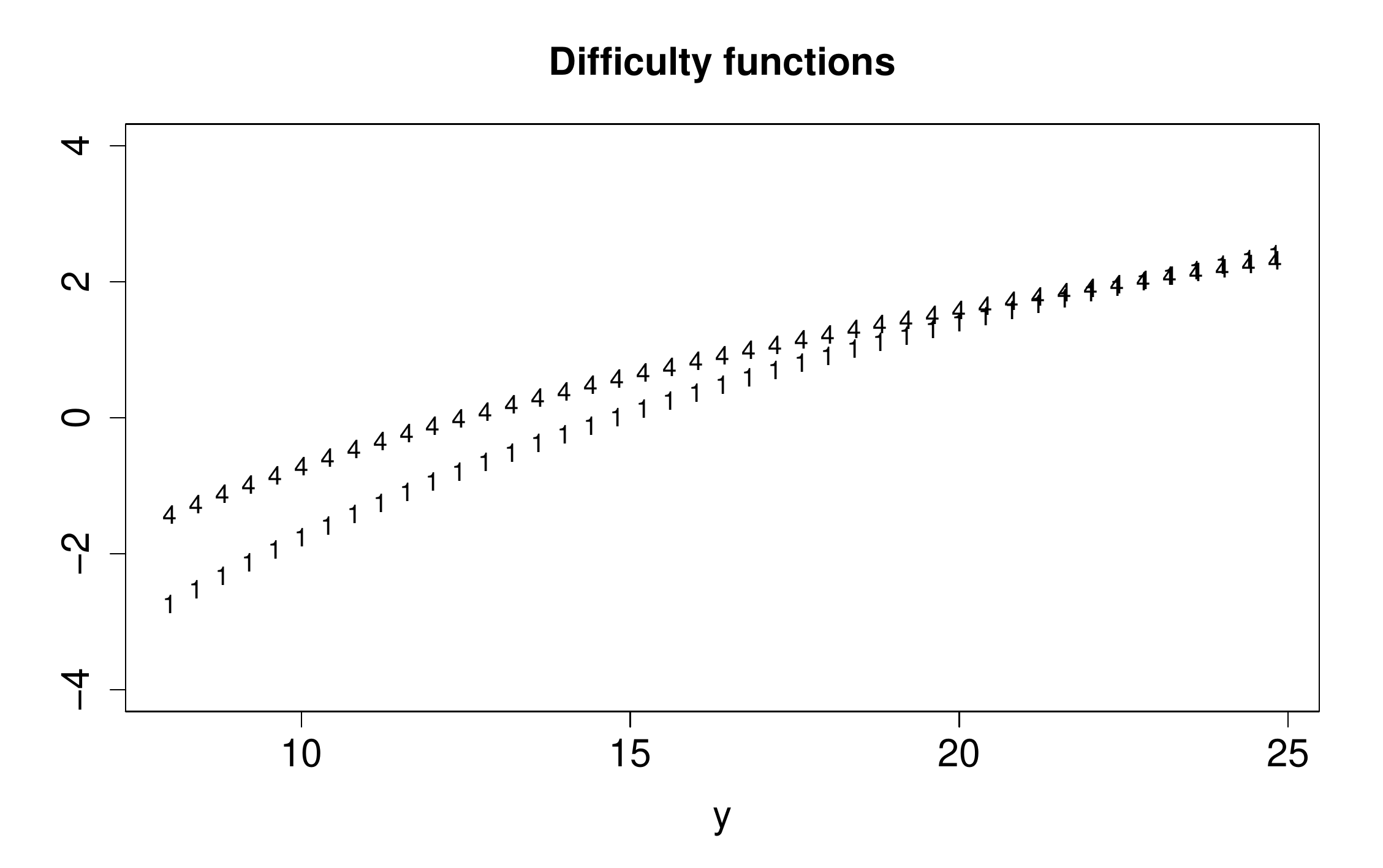}
\includegraphics[width=7cm]{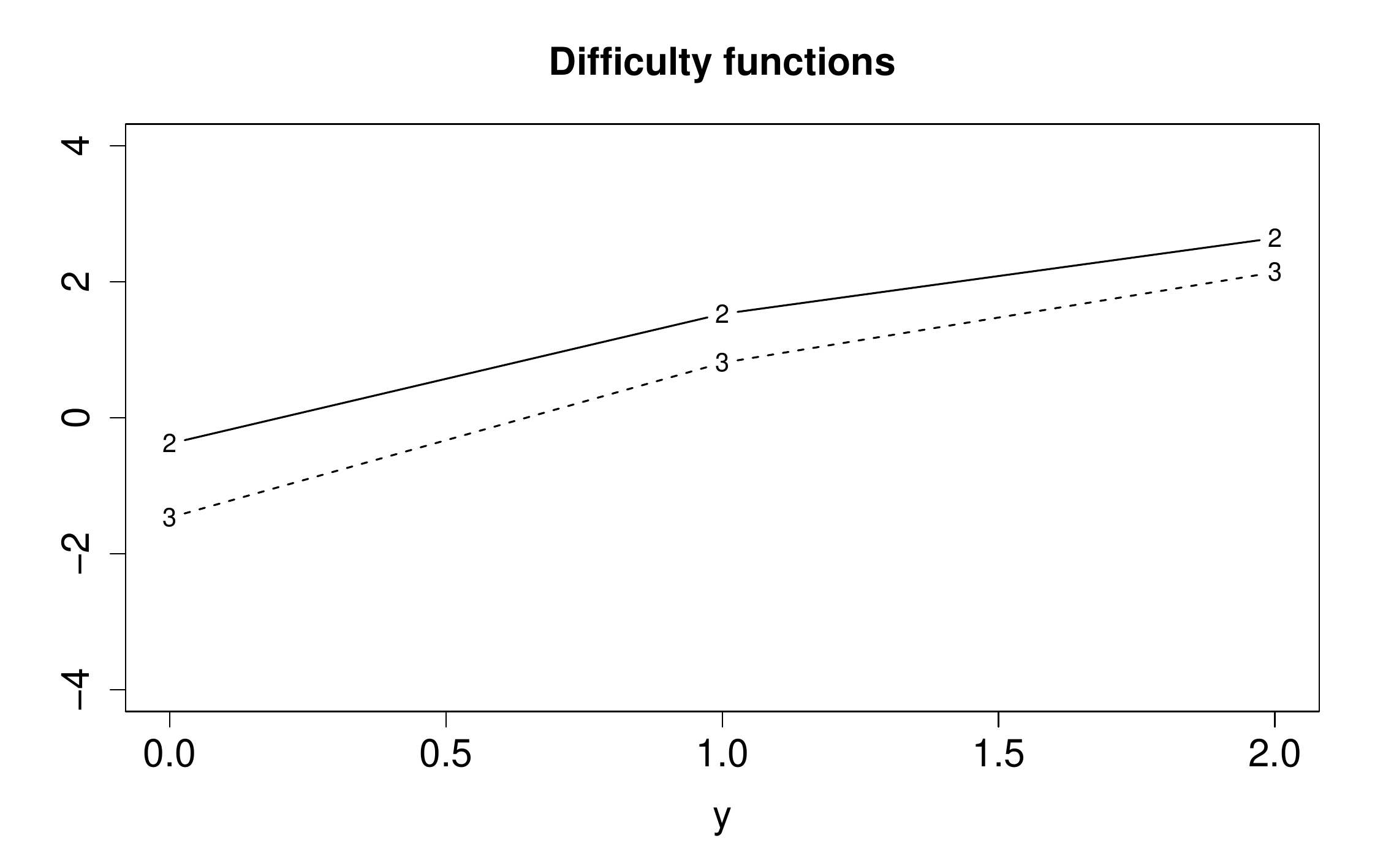}
\includegraphics[width=7cm]{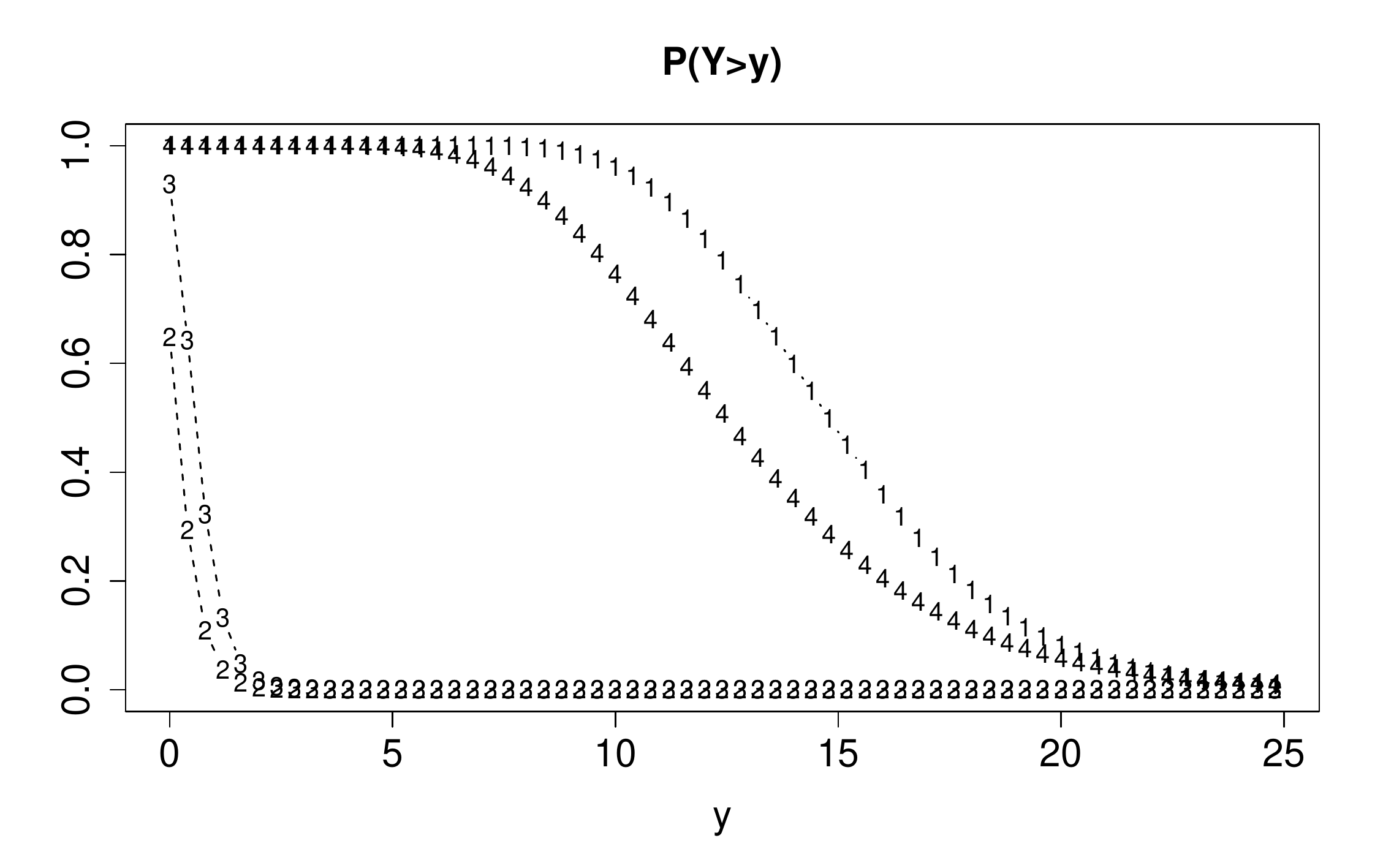}
\caption{First row: difficulty functions for fluency data with items 2 and 3 changed to three-categories items, below: person threshold functions for $\theta=0$.}
\label{fig:fluency3}
\end{figure}

A similar experiment was made for the verbal fluency data which have support $0,1,\dots$. Items 2 and 3 have been changed to three-categories items with support 0,1,2 by using thresholds 9 and 14. 
Figure \ref{fig:fluency3} shows the difficulty and PT functions for the mixed-formats case. As for the cognition data the PT functions for the unchanged items are very similar to the fits for the original items (see Figure \ref{fig:fluency11}, right picture) but the PT functions for the items 2 and 3 have distinctly changed since the new $y$-values are 0,1,2, which correspond to 0,9,14 on the original response scale.

\section{More General Models: Flexible Difficulty Functions}\label{sec:flexible}

The choice of the difficulty function   determines the response distribution beyond the choice of the response function. As shown before it can in particular be used to restrict the support of the response. A fixed choice,  for example  by using linear difficulty functions, assumes that items differ only by intercepts and slopes (of the difficulty function).  A fixed choice has the advantage that each item is determined by just two parameters $\delta_{0i}, \delta_i$,
a disadvantage is that the true difficulty function and  the distribution of the responses, which depends on the difficulty function, are typically unknown.

A more flexible approach that avoids that one has to choose a specific type of function, and lets the data themselves decide is obtained by letting    difficulty  
functions be determined by basis functions, an approach that has been extensively used   in statistics and machine learning \citep{Vidakovic:99,Wood:2006a,Wood:2006b,RupWanCar:2008,Wand:2000}. Let us  assume that the difficulty functions are given by
\begin{equation}\label{eq:diff}
\delta_{i}(y) = \sum_{l=0}^M \delta_{il}\Phi_{il}(y),
\end{equation}
where $\Phi_{il}(.), l =0,\dots, M$ are chosen basis functions. The simple choice $\Phi_{i0}(y)=1, \Phi_{i1}(y)=y, M=1$ means that the item difficulty functions are linear. Much more flexible models are obtained by alternative functions as radial basis functions or spline functions. A particular attractive choice are B-splines as propagated and motivated extensively by \citet{EilMar:96,eilers2021practical}. They are very flexible and can closely approximate a variety of functions. In the literature they were typically used to approximate functions of observable variables, here they are used to specify the unobservable difficulty functions. If difficulty functions are expanded in basis functions they have to fulfill that they are non-decreasing, which typically calls for some restrictions. In the case of B-splines a restriction that ensures that functions are non-decreasing is that $\delta_{i0} \le \dots \le \delta_{iM}$.

Although computation is more demanding, flexible difficulties can be used as a diagnostic tool to investigate if a fixed difficulty function is appropriate. It can also be used to investigate if single items have quite different distributions. One should distinguish between two cases, difficulty function as specified in equ. (\ref{eq:diff}), which vary freely across items, and  a slightly more restrictive approach, which assumes that only the location varies across items. The latter uses the simpler expansion 
\begin{equation}\label{eq:diffsimp}
\delta_{i}(y) = \delta_{0i}+ \sum_{l=0}^M \delta_{l}\Phi_{l}(y).
\end{equation}
It assumes that the location $\delta_{0i}$ is item-specific, but the form of the function is the same or all items. 

\begin{figure}[H]
\centering
\includegraphics[width=7cm]{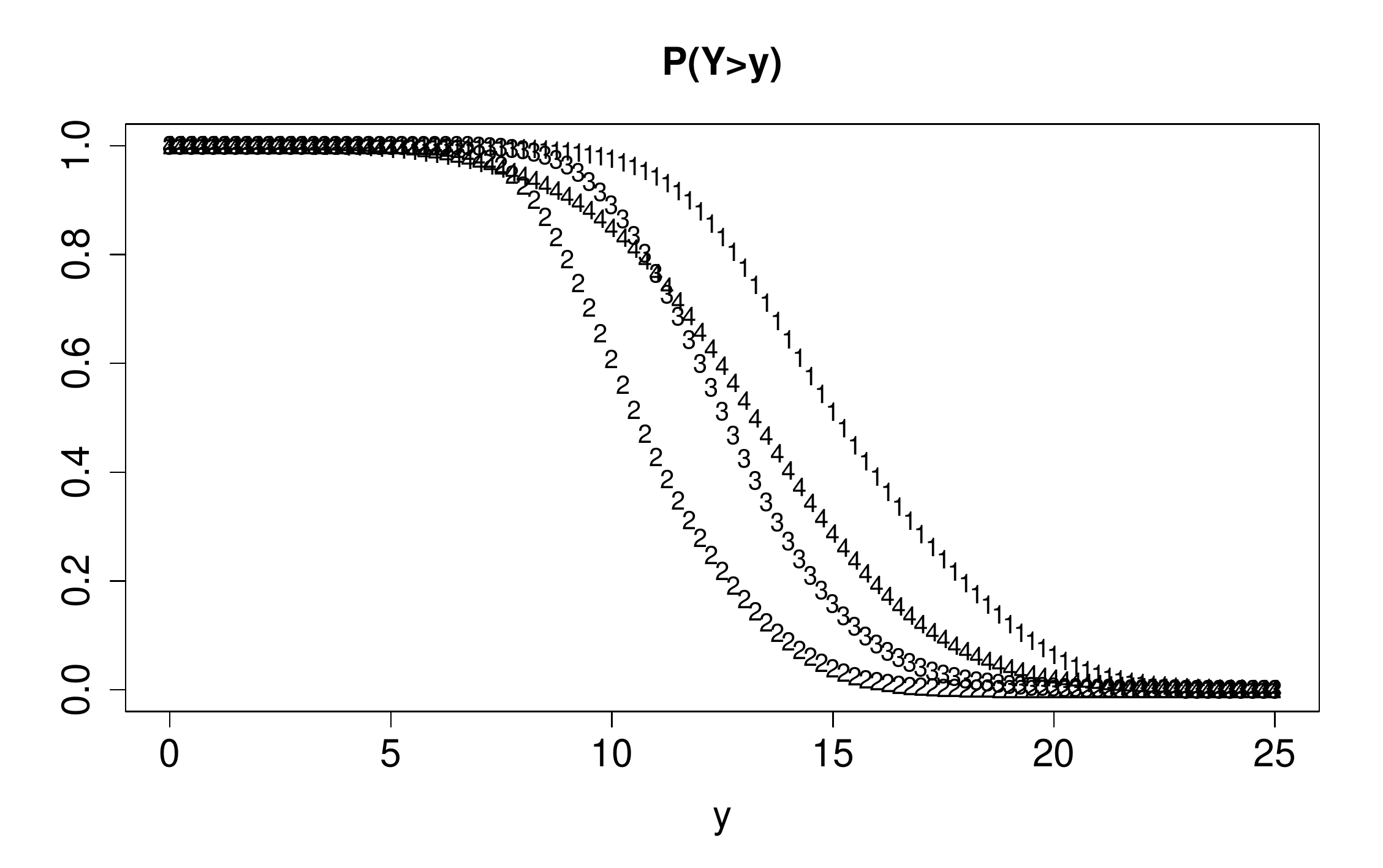}
\includegraphics[width=7cm]{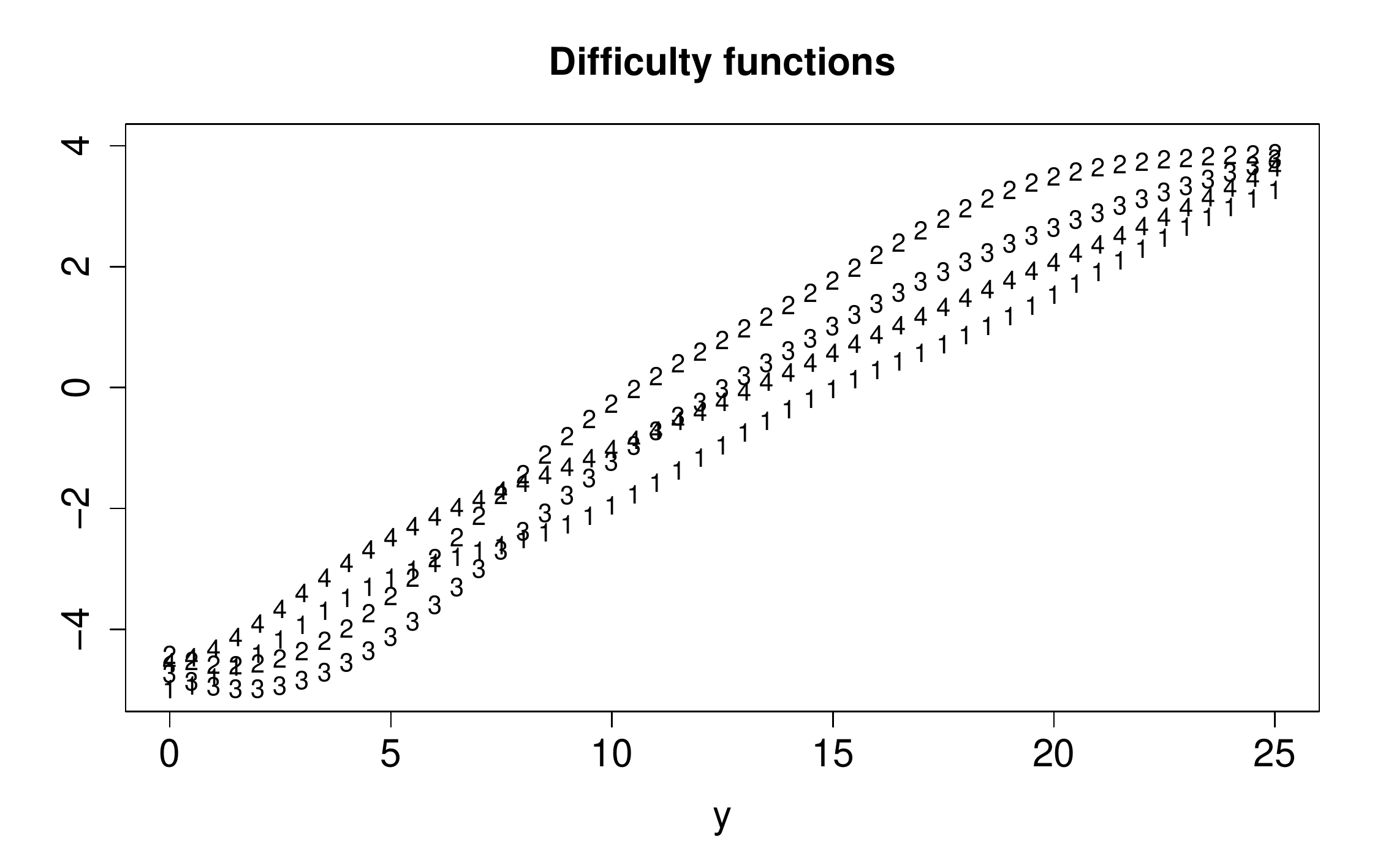}
\caption{Fitted PT (left) and difficulty (right) functions with  B-spline based difficulty functions for fluency data.}
\label{fig:fluency4}
\end{figure}

Figure \ref{fig:fluency4} shows the PT functions and the fitted difficulty functions  for the verbal fluency data if  difficulties are not restricted (8 cubic spline functions). It is seen that the obtained PT functions are very  similar to the functions obtained for  fixed logarithmic difficulty functions (Figure \ref{fig:fluency11}, right picture). The difficulty functions are close to linear function above $y=4$, which was the smallest observed value. In this range also the fitted logarithmic functions resemble linear functions. Therefore, there seems no substantial improvement over the model with logarithmic difficulty functions. This is also supported by the estimated standard deviation of person parameters, which are almost the same for both models ($\sigma_{\theta}=1.09$  for the splines model, $\sigma_{\theta}=1.11$ for the model with varying slopes). Of course,  the log-likelihood for the spline model is larger (loglik $= -1998.13$ for the splines model, -2038.58 for the model with varying slopes) but since the models are not nested differences of log-likelihoods are not informative. 


\begin{figure}[H]
\centering
\includegraphics[width=7cm]{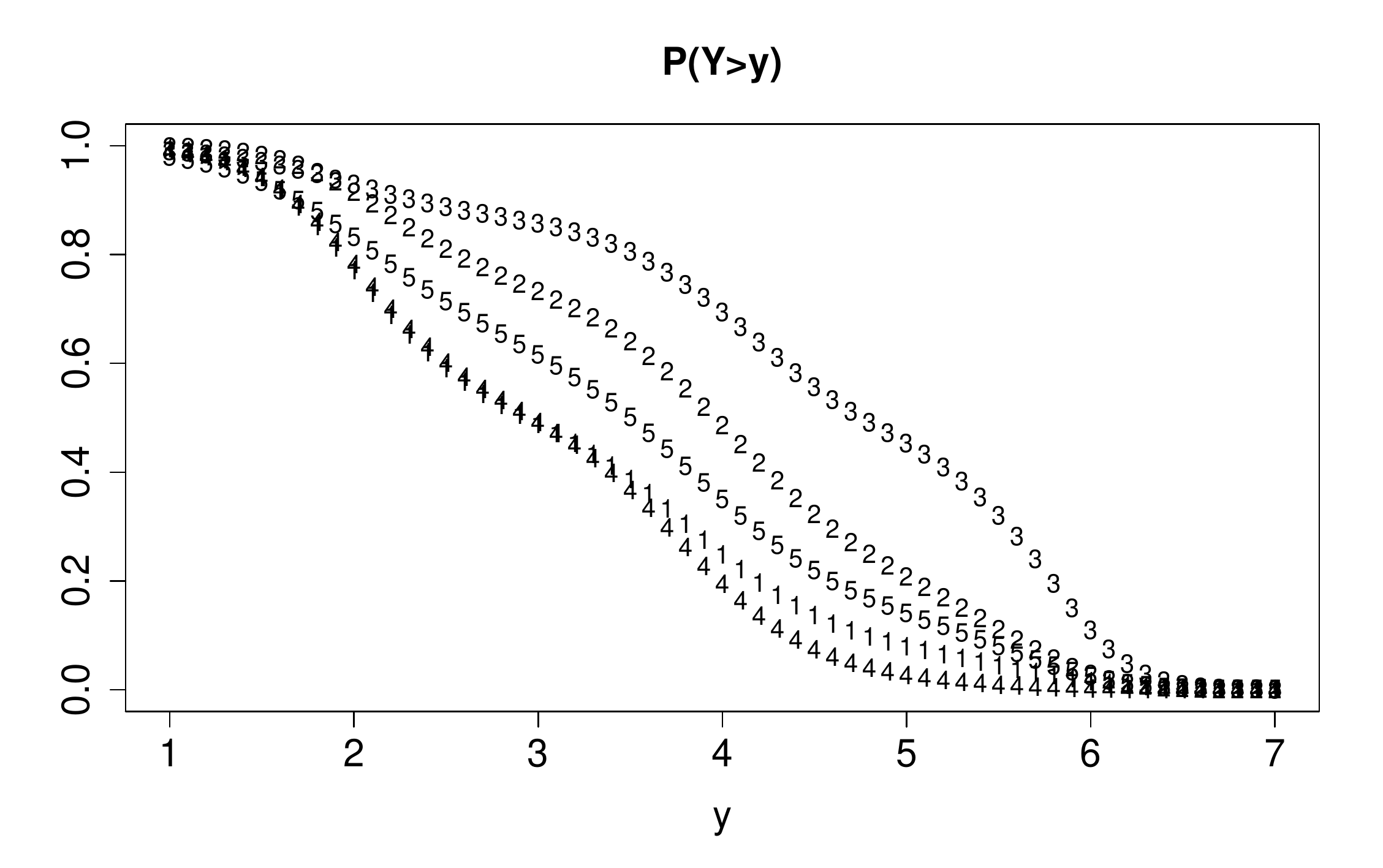}
\includegraphics[width=7cm]{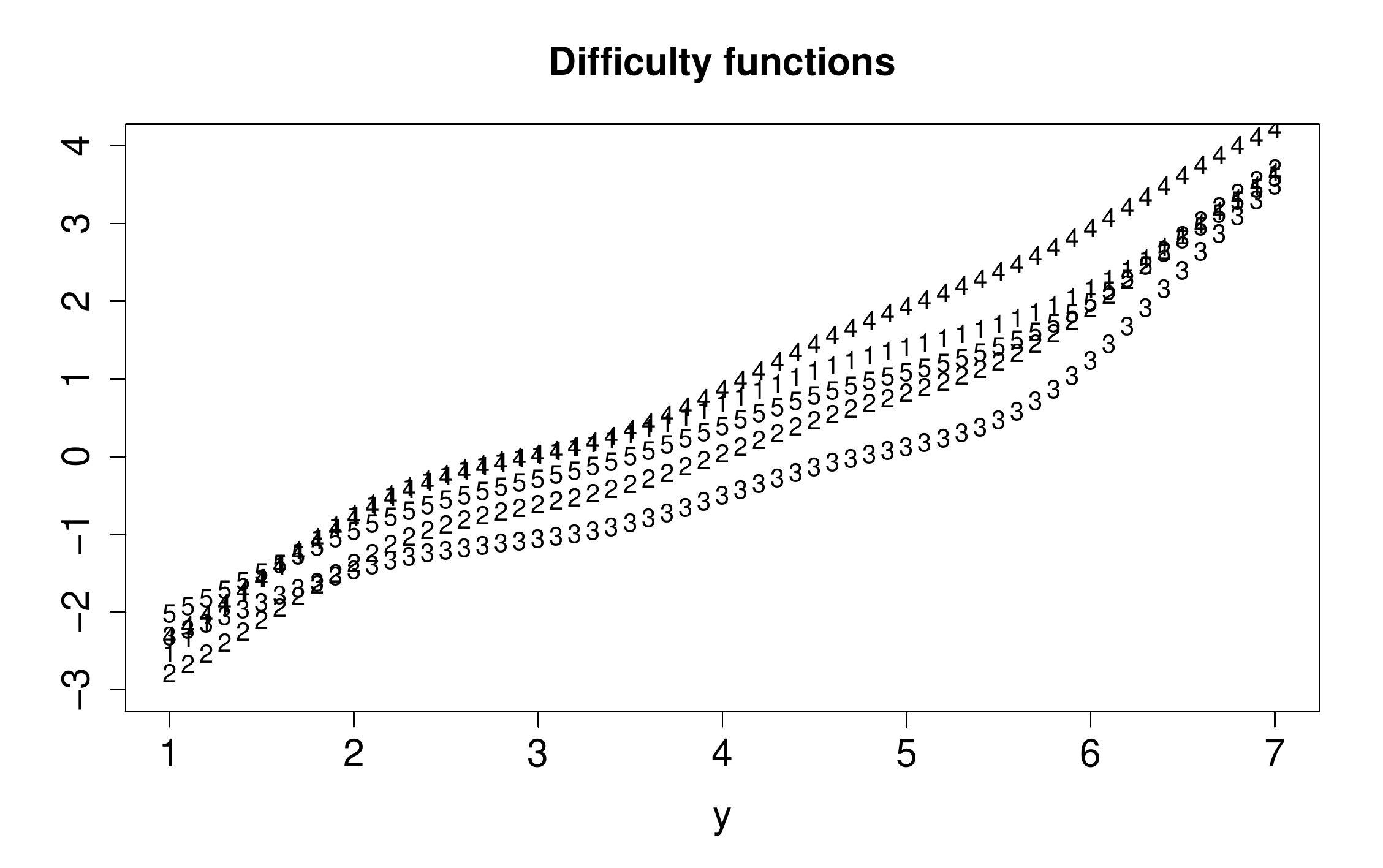}
\includegraphics[width=7cm]{fearsPTdiscr}
\includegraphics[width=7cm]{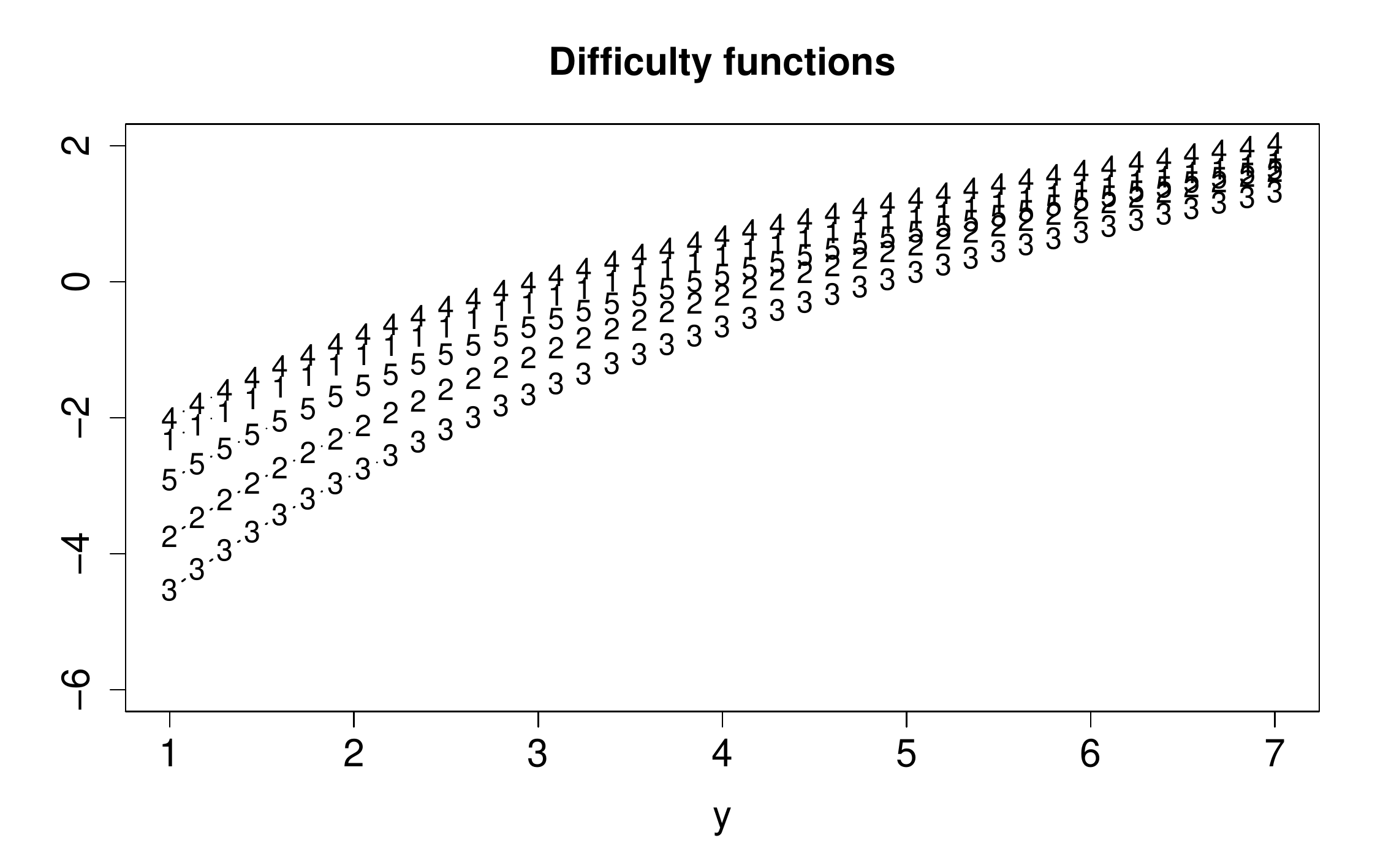}
\caption{Fitted PT (left) and difficulty (right) functions with  B-spline based difficulty functions for fear data; for comparison second row shows the difficulty functions for logarithmic difficulty functions}
\label{fig:fearsspl}
\end{figure}

Figure \ref{fig:fearsspl} shows the PT functions and the fitted difficulty functions  for the fear data if  B-splines (8 cubic spline functions) generate the difficulty functions, and a discrete distribution is used.
For comparison the second row shows the PT and difficulty functions for logarithmic difficulty functions. Though the order of the items remains the same, the form of the difficulty functions changes distinctly if splines are used  instead of the logarithmic function. Also  the   value of the log-likelihood (-1593.37) is much greater than the value obtained for logarithmic difficulty functions (-1987.26). 

\begin{figure}[H]
\centering
\includegraphics[width=7cm]{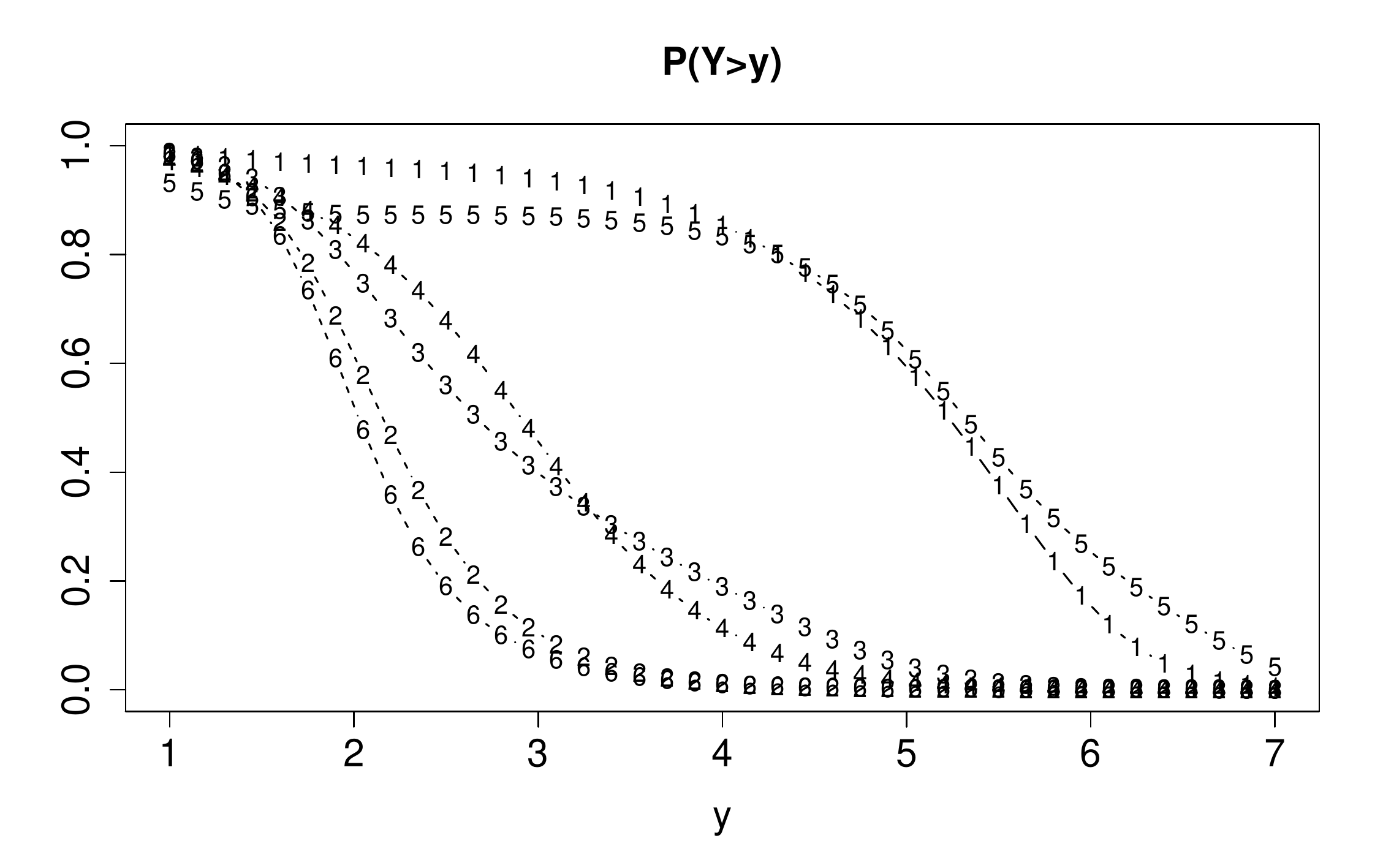}
\includegraphics[width=7cm]{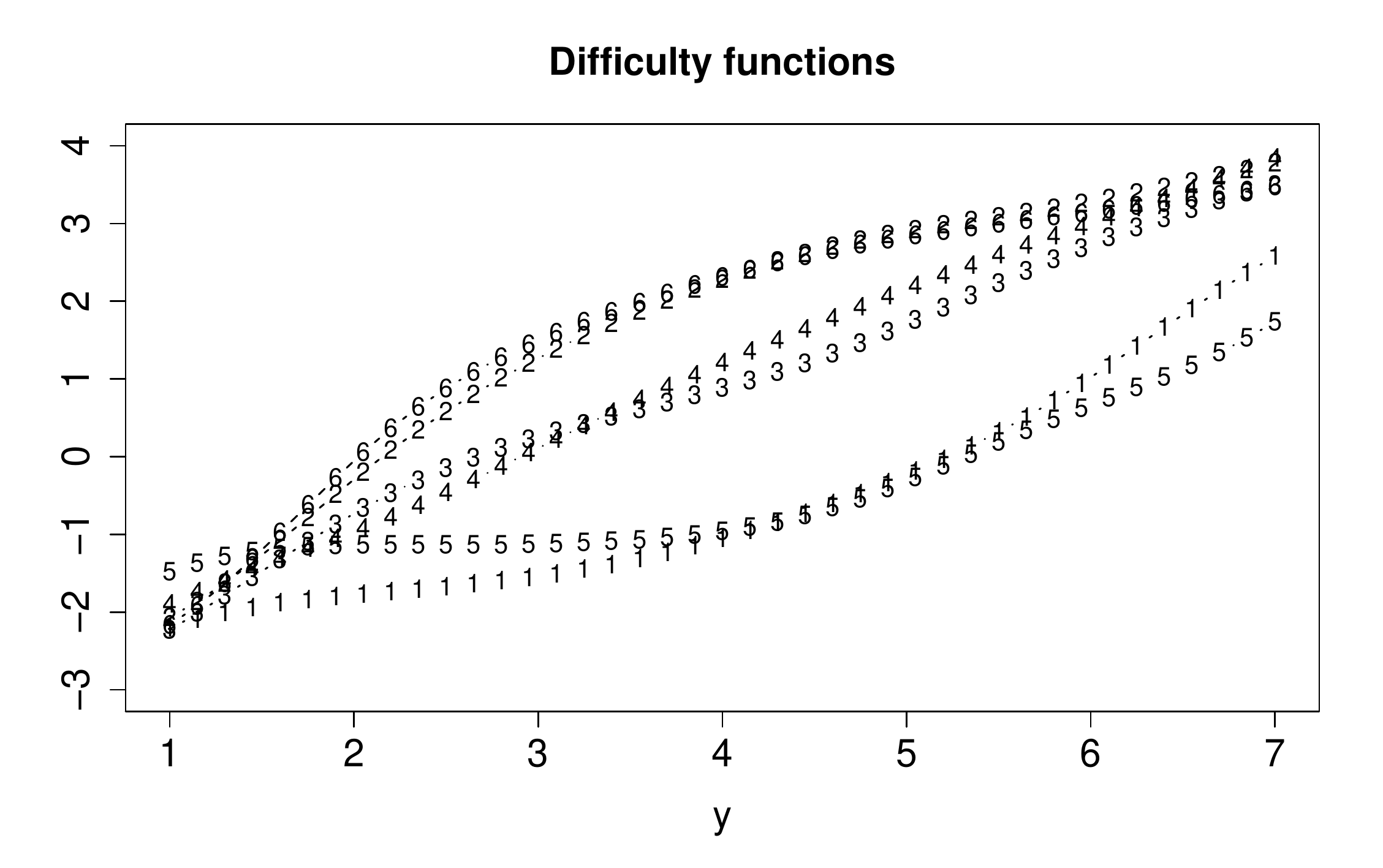}
\caption{Fitted PT (left) and difficulty (right) functions with  B-spline based difficulty functions for cognition data. }
\label{fig:lakes4}
\end{figure}

Figure \ref{fig:lakes4} shows the PT functions and the fitted difficulty functions  for the cognition data if  B-splines (8 cubic spline functions) generate the difficulty functions. It is seen that  difficulty functions differ from functions obtained for linear functions (see Figure \ref{fig:lakes1}) though the grouping in pairs of items is quite similar. It suggests that the response distributions deviate from the normal distribution, which is implicitly assumed by using linear difficulty functions.

\section{Obtaining Estimates and Inference }\label{sec:inference}
In the following marginal maximum likelihood methods for the estimation of item parameters and posterior estimation of person parameters are considered under the usual assumption of conditional independence of observable variables given the latent variables.

\subsection{Marginal Maximum Likelihood Estimation}

Let the general thresholds model hold. Then the  distribution function for observation $Y_{pi}$ has the form
\[
F_{pi}(y) =P(Y_{pi} \le y) =  1-F(\theta_p-\delta_{i}(y)). 
\]
For \textit{continuous} response one obtains the density by building derivatives yielding
\[
f_{pi}(y)=\frac{\partial F_{pi}(y)}{\partial y} =  f(\theta_p-\delta_{i}(y))\delta_{i}'(y),
\]
where $f(.)$ is the density corresponding to $F(.)$, and $\delta_{i}'(y)= \partial \delta_{i}(y)/\partial y$ is the derivative  of the threshold function.

For \textit{discrete} response $Y_{pi} \in \{0,1,\dots\}$ the probability mass function is  obtained by building differences. Then, one has the discrete density function
\begin{align*}
f_{pi}(0)&= 1-P(Y_{pi} > 0)=1-F(\theta_p-\delta_{i}(0)),\\
f_{pi}(r)&= P(Y_{pi} > r-1)- P(Y_{pi} > r)=F(\theta_p-\delta_{i}(r-1))-F(\theta_p-\delta_{i}(r)),  r=1,2,\dots
\end{align*}
where $\sum_r f_{pi}(r)=1$. 
For simple binary responses one obtains 
\begin{equation*}
f_{pi}(0)= 1-F(\theta_p-\delta_{i}(0))\quad 
f_{pi}(1)= F(\theta_p-\delta_{i}(0)),
\end{equation*}
where $\delta_{0i}=\delta_{i}(0)$ is the familiar difficulty parameter.

If difficulties are expanded in basis functions they have the form
\[
\delta_{i}(y) = \sum_{l=0}^M \delta_{il}\Phi_{il}(y)= \Phib_{i}(y)^T\deltab_{i},
\]
where $\Phib_{i}(y)^T=(\Phi_{i0}(y),\dots,\Phi_{iM}(y))$,  $\deltab_{i}^T=(\delta_{i0},\dots,\delta_{iM})$.
The corresponding derivative is given by
\[
\delta_{i}'(y) = \sum_{l=0}^M \delta_{il}\Phi_{il}'(y)= \Phib_{i}'(y)^T\deltab_{i},
\]
where $\Phib_{i}'(y)^T=(\Phi_{i0}'(y),\dots,\Phi_{iM}'(y))$ is the vector of derivatives of basis functions.

Let now observations be given by $y_{pi}, i=1,\dots,I, p=1,\dots,P$.
The estimation method that is used is marginal likelihood by assuming that person parameters are normally distributed,    
 $\theta_p\sim N({0}, \sigma_{\theta}^2)$.  
Maximization of the marginal log-likelihood can be obtained by integration techniques. We use
numerical integration by Gauss-Hermite integration methods. Early versions for univariate random effects date back to \citet{Hinde:82} and
\citet{AndAit:85}.

Let $\deltab_{i}$ denote the vector of all parameters linked to item $i$. For fixed difficulty functions it has length two, for expansions in basis functions it is, more generally,  $M+1$. With $\deltab ^T=(\deltab_{1}^T,\dots,\deltab_{I}^T,\sigma_{\theta})$ denoting  the set of all item parameters and $f_{0,\sigma_{\theta}}(.)$ denoting the density of the normal distribution $N({0}, \sigma_{\theta}^2)$ 
 the marginal likelihood has the form
\[
L(\deltab)=\prod_{p=1}^P \int  \prod_{i=1}^I f_{pi}(y_{pi}) f_{0,\sigma_{\theta}}(\theta_p) d\theta_p,
\]

yielding the log-likelihood
\[
l(\deltab)= \log (L(\deltab)) = \sum_{p=1}^P \log(\int  \prod_{i=1}^I f_{pi}(y_{pi}) f_{0,\sigma_{\theta}}(\theta_p) d\theta_p).
\]

The score function $s(\deltab) = \partial l/ \partial \deltab$, which is useful when computing estimates, has components
\begin{align*}
\frac{\partial l}{ \partial \delta_{ij}} &= \sum_{p=1}^P \int\frac{ \partial f_{pi}(y_{pi})}{ \partial \delta_{ij}} \prod_{l \ne i}f_{pi}(y_{pl})f_{0,\sigma_{\theta}}(\theta_p) d\theta_p / c_p,\\
\frac{\partial l}{ \partial \sigma_{\theta}}&= \sum_{p=1}^P \int\prod_{i=1}^I f_{pi}(y_{pi}) \frac{\partial f_{0,\sigma_{\theta}}(\theta_p)}{\partial \sigma_{\theta}} d\theta_p/ c_p.
\end{align*}
The form of the derivatives depends on the distribution of the responses. For continuous responses one obtains
\[
\frac{ \partial f_{pi}(y_{pi})}{ \partial \delta_{ij}} = \Phi_{ij}'(y_{pi})(f(\theta_p-\delta_{i}(y_{pi}))-f'(\theta_p-\delta_{i}(y_{pi}))\Phi_{ij}(y_{pi})\Phib_{i}'(y_{pi})^T\deltab_{i}),
\]
with $f'(.)$ denoting the derivative of $f(.)$, and $c_p = \int  \prod_{i=1}^I f_{pi}(y_{pi}) f_{0,\sigma_{\theta}}(\theta_p) d\theta_p$.  For discrete responses one has
\[
\frac{ \partial f_{pi}(y_{pi})}{ \partial \delta_{ij}} = -f(\theta_p-\delta_{i}(y_{pi}-1))\Phi_{ij}(y_{pi}-1)+ f(\theta_p-\delta_{i}(y_{pi}))\Phi_{ij}(y_{pi}),
\]
where $\Phi_{ij}(-1)$ is defined by $\Phi_{ij}(-1)=0$.

For simple difficulty functions the score functions simplify accordingly. For example, when the difficulty functions are linear one  has  
$\Phib_{i}(y)^T=(1,y)$, and $\Phib_{i}'(y)^T=(0,1)$.
An approximation of the covariance of the estimate, $\cov(\hat\deltab)$, is obtained by the observed information $-\partial^2 l/ \partial \deltab \partial \deltab^T$.

Some caution is needed when fitting the flexible model with difficulty functions determined by B-splines. Since B-splines sum up to 1 at any given value the parameters in the version with common difficulty function (\ref{eq:diffsimp}) are not identified. This can be fixed by choosing a fixed value for one of the parameters $\delta_{0i}$, for example, $\delta_{01}=0$. One can also use the general form (\ref{eq:diff}) and use a tailored penalty.
Instead of maximizing the log-likelihood one maximizes the penalized log-likelihood $l(\{\delta_i\})= l(\{\delta_i\}) -  P_{\lambda}(\{\delta_i\})$ with 
\[
P_{\lambda}(\{\delta_i\}) = \lambda\sum_{i=2}^I \sum_{l=2}^M [(\delta_{il}-\delta_{i,l-1})- (\delta_{i-1,l}-\delta_{i-1,l-1})]^2.
\]
For $\lambda\longrightarrow \infty$ the differences of adjacent parameters become the same for all items, such that the levels of the functions can differ but not the form of the function.

\subsection{Estimating Person Parameters}
\begin{figure}[H]
\centering
\includegraphics[width=5cm]{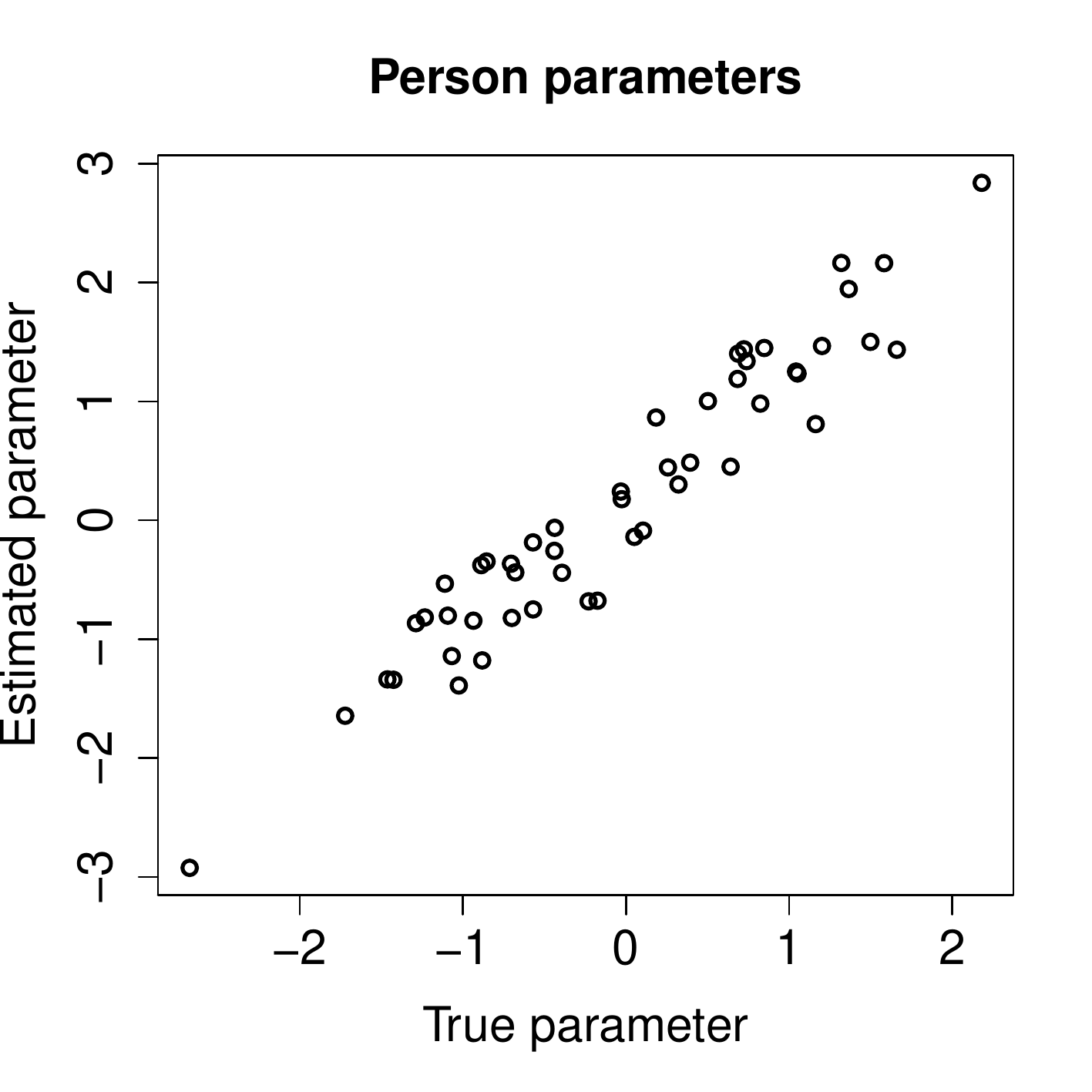}
\includegraphics[width=5cm]{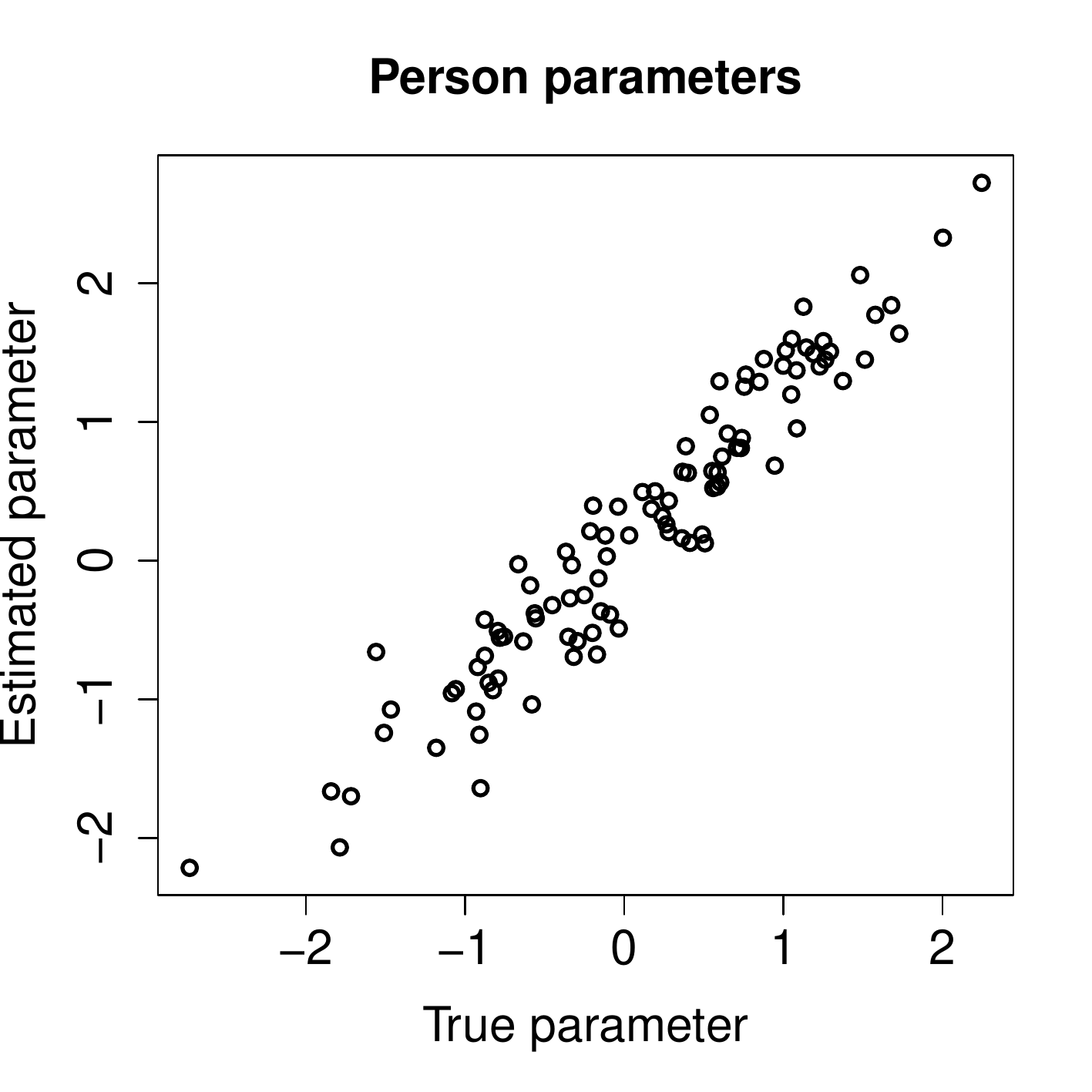}
\caption{True person parameters plotted against fitted values for simulation data for P=50 (left) and  P=100 (right).}
\label{fig:simpost}
\end{figure}

If estimates of item parameter are found, posterior mode or mean estimation yields estimates of person parameters. For given item responses
$\yb^T=(y_1,\dots,y_I)$, the posterior is given by 
\[
f(\theta|\yb, \deltab) = \frac{\prod_{i=1}^I f(\theta_p-\delta_{i}(y_i))\delta_{i}'(y_i) f_{0,\sigma_{\theta}}(\theta_p)} {\int  \prod_{i=1}^I f(\theta_p-\delta_{i}(y_i))\delta_{i}'(y_i) f_{0,\sigma_{\theta}}(\theta_p) d\theta_p}.
\]              
Replacing the parameter $\deltab$ by its estimate $\hat\deltab$ allows to compute the mode of the posterior $\hat\theta_m$ or the posterior mean
\[
\hat\theta_{m}=\E(\theta|\yb, \deltab) = \int \theta f(\theta|\yb, \deltab) d\theta.
\]
Figure \ref{fig:simpost} illustrates that posterior estimates are close to true values. Data were generated for 10 items with linear difficulty functions ($\delta_{i0}= -2.25 +(i-1)0.5$ $i=1,10$. For the first four items $\delta_{i}=1$, for the next four items $\delta_{i}=2$, and for the remaining two items  $\delta_{i}=3$. For $P= 50$ and $P=100$ with $\theta_p$ drawn randomly from $N({0}, 1)$ one obtains the plots of true person parameters against estimated parameters shown in Figure \ref{fig:simpost}.

\section{Concluding Remarks}
The comprehensive class of  thresholds models has been introduced and illustrated in examples. Also basic properties of the model class have been shown. 
Future research might be devoted to extensions of the model class and further investigations of its properties.
It is, for example, straightforward to include explanatory variables by using the additive term $\theta_p +\xb_p^T\betab- \delta_{i}(y)$ instead of the simple term $\theta_p -\delta_{i}(y)$, where $\xb_p$ is a person-specific explanatory variable and $\betab$ the corresponding weight. The latter can also be item-specific.
The incorporation of explanatory variables can be useful to investigate sources of heterogeneity in a response scales, and has been propagated, for example, by 
\citet{jeon2016generalized}. 
The models can also be extended to include a slope parameter by assuming the multiplicative term $\alpha_i\theta_p - \delta_{0i}$  with an item-specific slope.
The extension is not needed for continuous response with varying slopes in item difficulties since a slope parameter is already included. Also the consideration of multidimensional person parameters is a possible topic of further research.

We restricted consideration to symmetric response functions $F(.)$. The use of a normal or a logistic response function yields very similar results, although the scaling is different. However, the use of   non-symmetric distributions as, for example,  the extreme value distribution might make a difference. In principle also 
discrete response functions could be used, the extreme case being a zero-one function as in the Guttman model, however they include jumps that might be less realistic when assuming a continuous latent trait.   

Software for the computation of marginal maximum likelihood estimates will be made available on Github. 

\bibliography{literatur}

\section*{Appendix}
\begin{theorem} \label{try} 
If the item difficulty function in the thresholds model with continuous distribution function $F(.)$ and corresponding density $f(y) = \partial F(y)/\partial y$ is linear, $\delta_{i}(y)= \delta_{0i}+ \delta_i y$, $\delta_i \ge 0$ one obtains for the expectation and the variance
\begin{align}
&\E(Y_{pi}) = (\theta_p - \delta_{0i} - E_F)/\delta_i,\\
&\var(Y_{pi}) = var_F /\delta_i^2,
\end{align}
where $E_F=\int y f(y)dy$ is the expectation corresponding to distribution function $F(.)$, and $var_F =\int (y-E_F)^2f(y)d y$ is the variance linked to $F(.)$

In addition, the distribution function of $Y_{pi}$ is a shifted and scaled version of $F(.)$.
\end{theorem}

Proof:
For linear item function the thresholds model has the form $P(Y_{pi} > y|\theta_p,\delta_{i}(.))=F(\theta_p-\delta_{0i}- \delta_i y)$. The corresponding distribution function is
\[
F_{Y_{pi}}(y) = P(Y_{pi} \le y) = 1 - F(\theta_p-\delta_{0i}- \delta_i y).
\]
The density is given by
\[
f_{Y_{pi}}(y)=\frac{\partial F_{Y_{pi}}(y)}{\partial y} =  f(\theta_p-\delta_{0i}- \delta_i y)\delta_i,
\]
yielding the expectation

\[                  
\E(Y_{pi})=  \delta_i \int y f(\theta_p-\delta_{0i}- \delta_i y) dy.
\]
With $\eta = \theta_p-\delta_{0i}- \delta_i y$   and $d\eta/dy=- \delta_i$ one obtains                   
\[                  
\E(Y_{pi})=   -\int_{\infty}^{-\infty} \frac{\theta_p-\delta_{0i}-\eta}{\delta_i} f(\eta) d\eta = \frac{\theta_p-\delta_{0i}-E_F}{\delta_i}
\]
where $E_F=\int y f(y)dy$ is a parameter that depends on $F$  but not  on $i$.

The variance is given by
\begin{align*}                  
\var(Y_{pi})&=   \int (y-\frac{\theta_p-\delta_{0i}-E_F}{\delta_i})^2 f(\theta_p-\delta_{0i}- \delta_i y)\delta_i dy
= \int (\frac{\eta-E_F}{\delta_i})^2 f(\eta) d\eta \\
&= \var_F /{\delta_i}^2,
\end{align*}
where $\var_F=\int ({\eta-E_F})^2 f(\eta) d\eta$.

\medskip
\begin{theorem} \label{equ} \label{prop:id}
For the thresholds model with continuous response function $F(.)$ let the difficulty functions be defined on the support $S$ 
taking values from $\mathbb{R} \cup \{\infty \}$.

(1) For infinite support  person parameters and item functions are identifiable if one person parameter is fixed (e.g. $\theta_1=0$) or
 $\delta_i (y_0)$ is fixed for one item and one value $y_0 \in S$.
 
(2) For discrete and finite support $\{m_1,\dots, m_k\}$ the person parameters and item functions are identifiable if one person parameter is fixed  or  $\delta_i (y_0)$ is fixed for one item and one value $y_0 \in \{m_1,\dots, m_{k-1}\}$. For $m_k$ one has $\delta_i (m_k)=\infty$ for all items.
\end{theorem} 

Proof:
Since the response function is strictly increasing one obtains for two parameterizations $\theta_p,\delta_i (.)$ and $\tilde\theta_p,\tilde\delta_i (.)$ 
that $\theta_p- \delta_i(y) = \tilde\theta_p- \tilde\delta_i(y)$ holds for all values $y \in S$.

If one chooses $\theta_1=0$, and accordingly $\tilde\theta_1=0$ one obtains immediately $\delta_i(y)=\tilde\delta_i(y)$ for all values $y \in S$. Then, also
$\theta_p = \tilde\theta_p$ holds for all $\theta_p$.

Let us now choose $\delta_i (y_0)=0$ for one value $y_0 \in S$ for infinite support $S$. Then one obtains $\theta_p = \tilde\theta_p$ holds for all $\theta_p$, and therefore $\delta_i (y)=\tilde\delta_i (y)$ for all $y$.

If the support is finite one has to choose $\delta_i (y_0)=0$ for $y_0 \in \{m_1,\dots, m_{k-1}\}$. For $\delta_i (m_k)$ one always has $\delta_i (m_k)=\infty$
since $1=P(Y_{pi}>m_k)=F(\theta_p- \delta_i(m_k))$ has to hold for any $\theta_p$.

\medskip
\begin{theorem} \label{equ} 
The thresholds model $P(Y_{pi} > y)=F(\theta_p-\delta_{i}(y))$ holds for continuous response $Y_{pi}$ iff the graded response model holds for all categorizations 
\[
Y_{pi}^{(c)} =r \quad \Longleftrightarrow \quad  Y_{pi} \in (\tau_{r}, \tau_{r+1}],
\]
where $\tau_{1} < \dots < \tau_{k}$ are any ordered thresholds.
\end{theorem} 

Proof: 
Let the threshold model $P(Y_{pi} > y|\theta_p,\delta_{i}(.))=F(\theta_p-\delta_{i}(y))$ hold for continuous response $Y_{pi}$ for some increasing function $\delta_{i}(y)$. Let a categorized version be defined by 
\[
Y_{pi}^{(c)} =r \text{ if }  Y_{pi} \in (\tau_{r}, \tau_{r+1}].
\]
for any partition $\tau_{0}= -\infty < \tau_{1} < \dots < \tau_{k}$.

One obtains $P(Y_{pi} > \tau_{r}|\theta_p,\delta_{i}(.))=F(\theta_p-\delta_{i}(\tau_{r}))$ and therefore with $\delta_{ir}=\delta_{i}(\tau_{r})$
\begin{equation}\label{eq:part}
P(Y_{pi}^{(c)}  \ge r)  =F(\theta_p-\delta_{ir}),
\end{equation}
which is the graded response model for discrete response $Y_{pi}^{(c)}$.

Let now the discretized version (\ref{eq:part}) hold for all discretizations. Let us consider the discretization 
$\tau_{1} < \dots < \tau_{k}$ with response $Y_{pi}^{(c)}$ and parameters $\delta_{ir}$, and the discretization 
$\tau_{1}+\Delta < \dots < \tau_{k}$ with response $Y_{pi}^{(c_{\Delta})}$ and parameters $\delta_{ir}^{(c_{\Delta})}$, where $\Delta < \tau_{2}-\tau_{1}$.
Let the difficulty function be defined by $\delta_i(\tau_{1}) = \delta_{i1}$, $\delta_i(\tau_{1}+\Delta) = \delta_{i1}^{(c_{\Delta})}$ to obtain 
\[
P(Y_{pi} > \tau_{1})= F(\theta_p-\delta_i(\tau_{1})), \quad P(Y_{pi} > \tau_{1}+\Delta)= F(\theta_p-\delta_i(\tau_{1}+\Delta)),
\]
Since this holds for any values $\tau_{1}, \Delta$ one obtains the threshold model for continuous response $Y_{pi}$.

\section*{Acknowledgment}
I want to thank Pascal Jordan for various helpful suggestions and comments.

\end{document}